\documentclass[twocolumn]{aastex631}
\usepackage{xspace}
\usepackage{multirow}
\usepackage{amsmath}	
\usepackage{natbib}
\usepackage{mathtools}
\usepackage{comment}

\newcommand{\msun}{\mbox{M$_\odot$}}
\newcommand{\lsun}{\mbox{L$_\odot$}}
\newcommand{\erg}{\mbox{\rm erg}}
\newcommand{\yr}{\mbox{${\rm yr}$}}
\newcommand{\myr}{\mbox{${\rm Myr}$}}
\newcommand{\gyr}{\mbox{${\rm Gyr}$}}
\newcommand{\pc}{\mbox{${\rm pc}$}}

\newcommand{\kms}{\mbox{${\rm km}~{\rm s}^{-1}$}}

\newcommand{\feh}{\mbox{$[{\rm Fe}/{\rm H}]$}}

\newcommand{\hcc}{\mbox{${\rm H~cm^{-3}}$}}
\newcommand{\msunpc}{{\msun\,\pc^{-2}}}
\newcommand{\be}{\begin{equation}}
\newcommand{\ee}{\end{equation}}
\newcommand{\bea}{\begin{eqnarray}}
\newcommand{\eea}{\end{eqnarray}}

\newcommand{\code}[1]{\textsc{#1}}
\newcommand{\gizmo}{\code{GIZMO}\xspace}

\begin{document}

\title{The Star Clusters As Links between galaxy Evolution and Star formation (SCALES) project I: Numerical method}

\author[0000-0002-8556-4280]{Marta Reina-Campos}
\affiliation{Canadian Institute for Theoretical Astrophysics (CITA), University of Toronto, 60 St George St, Toronto, M5S 3H8, Canada}
\affiliation{Department of Physics \& Astronomy, McMaster University, 1280 Main Street West, Hamilton, L8S 4M1, Canada}

\author[0000-0001-9852-9954]{Oleg Y. Gnedin}
\affiliation{Department of Astronomy, University of Michigan, Ann Arbor, MI 48109, USA}

\author[0000-0003-3551-5090]{Alison Sills}
\affiliation{Department of Physics \& Astronomy, McMaster University, 1280 Main Street West, Hamilton, L8S 4M1, Canada}

\author[0000-0002-1253-2763]{Hui Li}
\affiliation{Department of Astronomy, Tsinghua University, Beijing 100084, People’s Republic of China}

\correspondingauthor{Marta Reina-Campos}
\email{reinacampos@cita.utoronto.ca}



\begin{abstract}
Stellar clusters are critical constituents within galaxies: they are the result of highest-density star formation, and through their spatially and temporally correlated feedback they regulate their host galaxy evolution. We present a novel numerical method to model star clusters as individual units of star formation using sink particles. In our method, star clusters grow via gas accretion and via merging with less massive clusters. We describe the implementation in the radiation hydrodynamics code \gizmo and run a large grid of marginally bound, turbulent clouds of $10^7~\msun$ to explore the effect of modeling ingredients on the evolution of the clouds and the star clusters. We find both gas accretion and mergers to be critical processes to form star clusters of masses up to $\sim10^5$--$10^6~\msun$, while ionising radiation is the main feedback mechanism regulating the growth of star clusters. The majority of our star clusters assemble their mass in $0.3$--$2.6~\myr$, and the most massive ones take $\sim10~\myr$. By removing high density gas by accretion, our sink-based cluster formation prescription allows the newly-formed star clusters to inject their stellar feedback in less dense environments. This makes feedback more efficient at ionising and disrupting the cloud than if we were to use a standard star formation approach, indicating that our numerical method is the missing critical step to model the interplay between star clusters and their host galaxies.
\end{abstract}



\section{Introduction} \label{sec:intro}

Star clusters provide a critical link between star formation and galaxy evolution \citep[e.g.][]{forbes18}. The majority of stars form in spatially clustered environments \citep{elmegreen00, lada03} that can hierarchically assemble into gravitationally-bound clusters \citep[e.g.][]{howard18, dobbs22, dellacroce23}. The properties of the nascent star cluster population, together with their evolution, are heavily influenced by the gas conditions in which they form \citep[e.g.][]{krumholz19,adamo20}.

The clustering of star formation implies that the majority of massive stars release their feedback concentrated both in space and time \citep{keller15,fielding17,fielding18,orr22a,orr22b}. By  concurrently ejecting mass, momentum, energy and radiation into the interstellar medium, the clustered massive stars begin disrupting their natal environment and ionise their surroundings. This allows the clustered supernovae to drive powerful superbubbles capable of launching strong galactic outflows \citep{fielding18, sirresi24}. 

The intertwined evolution of star clusters and their host galaxies starts early in the lifetime of the Universe. Observations from the \textit{Hubble Space Telescope} and the \textit{James Webb Space Telescope} are finding ubiquitous nascent star clusters across cosmic time \citep[e.g.][]{vanzella17,vanzella22a,vanzella22b,vanzella23,adamo24, mowla24}, with the earliest a mere $460~\myr$ after the Big Bang \citep[$z=10.2$,][]{adamo24, bradley24}. These observations indicate that clustered stellar feedback can alter the evolution of galaxies at different epochs in their lifetimes. In the Local Universe, superbubbles driven by clustered winds and supernovae (SNe) produce recognisable features in the morphology of the interstellar medium (ISM) and redistribute the energy and turbulence back to the ISM \citep{watkins23}. These type of superbubbles are more likely to break out from the gas-rich, turbulent high-redshift galaxies \citep[$z\gtrsim2$;][]{fielding18,orr22b}, creating channels for the leakage of ionising radiation into the circumgalactic medium \citep{mcleod21}. The leaked radiation could contribute to the process of reionising the early Universe, for which star clusters have long been suggested as possible sources \citep[e.g.][]{ricotti02, griffen10, katz14, vanzella17, he20, vanzella20}. Similarly, the efficient expansion of superbubbles driven by massive star clusters has been posited as a possible formation pathway for ultra-diffuse galaxies \citep{trujillo-gomez22}.

In order to understand the impact of clustered feedback on the evolution of galaxies, we need numerical simulations capable of following the formation and evolution of galaxies and their star cluster populations over cosmic time. Previous efforts have focused on modelling this concurrent evolution using a sub-grid approach for the star clusters in cosmological zoom-in simulations of galaxy formation (e.g.~E-MOSAICS, \citealt{pfeffer18}, and EMP-\textit{Pathfinder}, \citealt{reina-campos22b}). Despite being able to capture the entire lifetime of the Universe for a large number of systems, this approach does not allow the star clusters to modify their galaxies. In contrast, current efforts focus on modelling the formation of star clusters with high-resolution single-star simulations of dwarf galaxies (e.g.~the GRIFFIN project, \citealt{lahen20}, the INFERNO simulations, \citealt{andersson23}, and the RIGEL project, \citealt{deng24}). In this approach, star clusters are bound groups of stellar particles of $m\sim\msun$, and the stellar feedback is tied to the evolution of the single-star particles. However, although in this method star clusters interact with their galactic environment, these simulations are severely limited in their time evolution and to modelling only dwarf galaxies due to huge computational costs. 

Therefore, we require a numerical prescription capable of bridging the scales between these approaches. One successful method seeds star clusters in dense gas regions, that would then accrete gas from their surrounding environment until feedback from the newly formed cluster terminates it \citep{li17,li18,li19a,brown22,bieri23}. However, this method neglects the relevance that sub-clusters mergers have on the growth of star clusters, which can account for about half their mass \citep{howard18}. In the method presented in this work, star clusters are modelled with sink particles that can grow over time via gas accretion and hierarchical merging, thus including the effect of their feedback on their host environment but without needing to add their internal collisional stellar dynamics. Thus, our prescription is suited for modeling a wide range of galactic environments over the lifetime of the Universe. 

In this paper, we introduce the numerical prescription for modelling individual star clusters with sink particles, and explore how different choices affect the formation of clusters and the evolution of isolated spherical clouds. The paper is organised as follows. We describe the numerical methods and our prescription in Sect.~\ref{sec:methods}. We outline the initial conditions and the numerical tests in Sect.~\ref{sec:ics-tests}. Sect.~\ref{sec:results} contains the results of our runs. A discussion of previous works and the caveats of the method is presented in Sect.~\ref{sec:discussion}, and the conclusions are presented in Sect.~\ref{sec:conclusions}. We include in App.~\ref{app:fb} the relevant equations used to model the different stellar feedback mechanisms from \citet{hopkins23}, and we describe in App.~\ref{app:numerical-tests} various numerical tests of our implementation.

\section{Numerical methods}\label{sec:methods}

\subsection{The hydrodynamical code GIZMO}

We use a public version of radiation hydrodynamics code GIZMO\footnote{Available at \href{https://bitbucket.org/phopkins/gizmo-public/}{https://bitbucket.org/phopkins/gizmo-public/}.} \citep{hopkins15}. It uses the Lagrangian Meshless Finite Mass method to solve the equations of hydrodynamics, and the M1 (first moment) approximation for the propagation of
radiation \citep{levermore84,rosdahl13,hopkins19}. This explicit method requires a Courant-Friedrichs-Lewy timestep criterion that imposes a minimum time required for light to traverse the cell. To reduce the computing cost associated with very small timesteps, we set a reduced speed of light $\tilde{c} = 0.01 c$. 

We use the most complete description of the cooling and heating physics available (described in app.~B in \citealt{hopkins18b} and updated in \citealt{hopkins23}), which extends the cooling curves down to $T\sim10~$K. These include collisional excitation and ionization, recombination, free-free emission, high- and low-temperature metal-line cooling, as well as fine-structure and molecular cooling; dust collisional heating and cooling, Compton heating and cooling; photoionization, cosmic ray and photoelectric heating; and optically thick cooling. For the metal-dependent processes, we use the full multi-species-dependent metal-line cooling tables from \code{CLOUDY} runs following \citet{wiersma09}. We use the simplest molecular fraction model, in which the fraction solely depends on the density and temperature of the gas \citep{glover12}, and the dust-to-metals ratio is a smooth function based on the dust temperature.  

\subsection{Star clusters as individual units of star formation}

The novel feature of our simulations is that the individual units of star formation are modelled as star clusters via sink particles. In our prescription, sinks are initially gas-rich, dense clumps of gas, which can grow in mass via gas accretion and hierarchical merging with gravitationally-bound sub-clusters \citep[e.g.][]{howard18}. Once their mass surpasses a threshold, stellar populations can form, and will continue to form as the sinks keep growing in mass.
The feedback output of the newly formed cluster is tied to the combined output of the stellar populations.

Since the process of sink formation is numerically equivalent to that of forming particles that represent single stellar populations (i.e. `standard star formation'), we describe it here interchangeably. We note that for both the formation of sinks out of the gas particles, and the formation of stellar populations within the sinks out of accreted gas, we use a formation efficiency of $100~$per cent. Here we describe the criteria to form sinks from the gas, and how star clusters can grow in mass by gas accretion and hierarchical merging with other sinks.

\subsubsection{Formation of star clusters via sinks}

Gas particles are required to meet a list of criteria to be eligible for sink formation:
\begin{itemize}
    \item \textit{Sufficiently dense}: the number density of hydrogen in the gas particle must exceed a threshold, $n_{\rm H} = \rho/(m_{\rm p}X) \geq n_{\rm th} = 10^4~\hcc$, which is resolution-dependent (see Sect.~\ref{sec:ics-tests}). To calculate it, we use the mass density of the particle, $\rho$, the proton mass $m_{\rm p} = 1.67\times10^{-24}~\rm g$, and $X = 0.76$ as the Hydrogen mass fraction. 
    \item \textit{Self-gravitating}: sink formation is allowed in gas particles with virial parameter $\alpha_{\rm vir}<1$. From \citet{hopkins13b}, the virial parameter is calculated on the scale of the cell, 
    \be
    \alpha_{i} = \dfrac{{{||\nabla\otimes \mathbf{v}||}^2}_{i} + {(c_{s, i}/h_{i})}^2}{8\pi G\rho_{i}},
    \ee
    where ${{||\nabla\otimes \mathbf{v}||}^2}_{i}$ is the Frobenius norm of the velocity gradient tensor, $c_{s, i}$ is the sound speed, and $h_{i}$ is the interparticle spacing of the particle $i$. This is the most restrictive criterion to trigger sink formation.
    \item \textit{Converging flows}: sink formation is only allowed in gas particles whose divergence of the velocity field is negative ($\nabla\cdot \mathbf{v} < 0$), indicating that the convergent flows will increase its density.
\end{itemize}

Once gas particles are flagged as sink-forming, we calculate their expected sink formation rate as
\be
\dot{m}_{\star} = \epsilon_{\rm ff} m_{\rm gas}/t_{\rm ff}.
\ee
We assume a constant star formation efficiency per free-fall timescale of $\epsilon_{\rm ff} = 100~$per cent \citep[following][]{hopkins18b,li19b,grudic21}, and the free-fall timescale is $t_{\rm ff} = \sqrt{3\pi/(32G\rho)}$. 

The conversion of gas into sinks is a stochastic process, regulated by the sink formation rate and the mass of gas particles. For each sink-forming gas particle $i$, we calculate the probability of forming the expected stellar mass during timestep $\Delta t_{i}$ as
\begin{equation}
    p_{i} = \left[1-\exp\left(-\dfrac{\dot{m_{\star}}^i\Delta t_i}{m_{\rm gas}^i}\right)\right],
\end{equation}
and a sink is formed whenever this probability is larger than a random uniformly-drawn number between 0 and 1. The sink inherits its properties (i.e.~position, velocity, chemical abundances) from the gas particle.

\subsubsection{Mass growth via gas accretion and merging}

\begin{figure*}
    \centering{
	\includegraphics[width=\hsize]{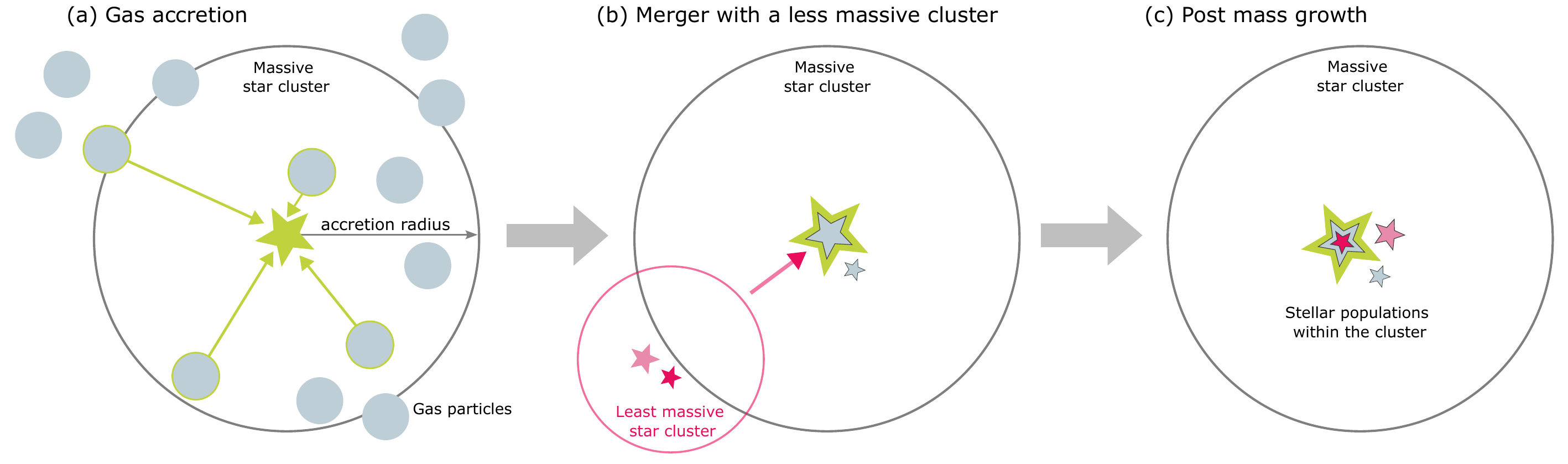}}
  \caption{Star clusters grow via accretion of gas particles (\textit{panel a}) and hierarchical merging with less massive clusters (\textit{panel b}). Depending on the locally-determined Bondi-Hoyle-Lyttleton accretion rate, gas particles within the accretion radius are tagged to be accreted (circles with green edge colors). Additionally, less massive clusters that are gravitationally bound to the massive cluster are merged into it (magenta circle).  The mass growth from accretion and merging can lead to a variety of outcomes (\textit{panel c}): if the accreted material is sufficiently distinct in metallicity from any of the current stellar populations, a new population forms within the cluster (blue star); otherwise, it contributes to one of the existing populations (blue star within the green star). Similarly, if the stellar populations coming from the less massive cluster are sufficiently distinct, they will be added as new (light magenta star); otherwise they will be absorbed into an existing population (magenta star within the green star). The masses, ages, and metallicities of the stellar populations determine the amount of mass, yields, energy and radiation emitted from the star cluster.}
  \label{fig:scheme-growth-sinks}
\end{figure*}

The mass of the sink particles can grow via two different mechanisms: they can stochastically accrete gas according to the Bondi-Hoyle rate, and they can merge hierarchically with less-massive bound sinks within their accretion volume. 

To calculate the feeding of each sink, we iterate over all the neighbours contained within the accretion radius of the sink, or whose volume of influence intersects with that of the sink. The accretion radius is set by the smoothing length of the sink, which is determined as the length scale enclosing the particle number density of 64 neighbours for a given kernel. The volume of influence of the neighbouring particle is defined as the largest between their smoothing length and gravitational softening. From these neighbours, we estimate the density of gas within the accretion region around the sink, $\rho_{\rm sink}$, the sound speed within the kernel, $c_{s,\rm sink}$, and the relative velocity between the gas in the kernel and the sink. 

With these quantities, we then relate the (unresolved) accretion onto the sink to the large-scale gas distribution using a Bondi-Hoyle-Lyttleton rate \citep{hoyle39, bondi44,bondi52},
\begin{equation}
    \dot{m} = \dfrac{4 \pi G^2 \, m_{\rm sink, exp}^2 \, \rho_{\rm sink}}{(c_{s, \rm sink}^2 +  |\sum_i (\mathbf{v}_{{\rm ngb}, i} - \mathbf{v}_{\rm sink})|^2)^{3/2}},\label{eq:bondi-hoyle-mdot}
\end{equation}
where $m_{\rm sink, exp} (t)$ is the expected mass of the sink assuming accretion happened smoothly rather than in quanta of $m_{\rm gas}$. As shown below, the inverse cube dependence on the sound speed makes this rate highly sensitive to the temperature of the gas, with higher gas temperatures leading to smaller accretion rates.

Given the accretion rate $\dot{m}$, the actual amount of gas accreted each timestep is determined stochastically. At a time $t$, the mass left to be accreted during that timestep, $\Delta m_{\rm to~acc}$, is the difference between the expected mass of the sink and the cumulative amount of gas flagged to be accreted during that timestep. Looping over the gas neighbours within the accretion region of the sink, each of them is assigned a probability to be accreted, $p_i = (\Delta m_{\rm to~acc} w_{i})/\rho_{\rm sink}$, where $w_i = W(\mathbf{x}_i-\mathbf{x}_{\rm sink})$ is the density of the neighbour on the kernel of the sink. This probability tends to zero as the cumulative amount of gas flagged to be accreted during that timestep reaches the expected amount. If the probability to be accreted is larger than a randomly-drawn number uniformly distributed between $0$--$1$, the gas neighbour is tagged for accretion, and the expected mass to be accreted is reduced accordingly.

Using a stochastic process for modelling gas accretion has both perks and caveats. By randomly picking gas neighbours based on the accretion rate, the prescription is agnostic to the properties and the geometry of the gas being flagged, i.e. it does not require gas to be dense or in certain structures to be accreted. Thus, the prescription only relies on the resolved large-scale gas distribution and is insensitive to the unresolved small-scale filamentary structure of the interstellar medium (ISM). However, there are two limiting cases in which this rate of accretion can lead to un-physical results. First, the accretion rate can diverge if there are high density peaks with close to null relative velocity within the kernel of the sink. Alternatively, the predicted mass to be accreted within a timestep $\dot{m}\Delta t$ can become larger than the actual mass within the kernel of accretion of the sink. To avoid both of these scenarios, we impose an upper limit on the predicted mass to be accreted to be less than the mass within the kernel.

In addition to accreting gas, sinks can also grow by merging with less-massive sinks. Every time a sink moves within the accretion region of another, we evaluate whether their relative velocity is less than that needed to escape, which is set to be at least $10~\kms$, and whether they are gravitationally bound. The criteria for boundedness requires that the largest distance the least massive sink can move assuming a purely radial orbit has to lie within twice the gravitational softening of the massive one. If both of these conditions are met, the more massive sink swallows the least massive one, hence dubbing this a `hierarchical' process. 

\subsection{Distinct stellar populations within star clusters}

As sinks grow mass via gas accretion and hierarchical merging, their gas reservoir also increases. Whenever the reservoir exceeds $300\,\msun$, we attempt to form a stellar population within the sink. For that, we check whether the new stellar population would have an age and metallicity within $\Delta \tau \leq 0.5~\myr$ and $\Delta [Z/H] \leq 0.05~\rm dex$ of any pre-existing stellar population of the sink. If so, we add up their masses and combine the ages and metallicities with a mass-weighting scheme. If either of the properties are more distinct than that, we form a new stellar population in the sink. The moment a sink forms a stellar population, we consider that it now represents a stellar cluster. 

Similarly, when two sinks merge, the stellar populations of the least massive sink are combined with the ones of the massive particle. If any of the merging stellar populations are closer in age than $\Delta \tau \leq 1~\myr$\footnote{Typical age uncertainties in young star clusters are of a few Myr \citep[e.g.][]{whitmore21}.} and with a metallicity difference less than $\Delta [Z/H] \leq 0.05~\rm dex$ from an existing one in the host, then the populations are combined as described above. Otherwise, the accreted stellar populations are added to the list of populations within the feeding sink. 

The properties of the stellar populations are given by their initial and current masses, age, total metallicity, as well as nine chemical abundances that track the enrichment from the different feedback mechanisms (He, C, N, O, Ne, Mg, Si, S, Ca, Fe; see Sect.~\ref{sub:fb}). Although we allow for up to 20 stellar populations per sink, the majority of clusters in our tests host between 1 and 9 populations. Keeping track of these individual distinct stellar populations allows us to measure age and metallicity spreads within the resulting star clusters.

\subsection{Stellar feedback}\label{sub:fb}

Stellar populations within the star clusters evolve independently from one another. To model their evolution, we consider continuous mass loss from OB and AGB winds, emission of radiation in five different bands (ionizing, far UV, near UV, optical and mid-IR, and far IR), as well as the energy and mass ejected by core-collapse (CC) and Ia supernovae (SNe). For this, we implement the prescriptions presented for FIRE-3 in \citet{hopkins23}. These feedback prescriptions are derived using STARBURST99 \citep{leitherer14} and assuming a three-part \citet{kroupa01} IMF, an $8~\msun$ SNe cut-off, the `evolution' wind model for O/B and `empirical' model for AGB mass-loss, and using the Geneva 2013 rotating stellar isochrone models \citep[e.g.][]{ekstrom12}. A key assumption of these prescriptions is that the IMF of the stellar populations producing the stellar feedback is well sampled, which imposes a minimum mass for the stellar populations within our star clusters that we set to $300~\msun$ \citep[see fig.4 in][]{smith21a}. 

We provide the relevant equations from \citet{hopkins23} implemented in our code in Appendix~\ref{app:fb}, and we refer the reader to the reference paper for more details. We use the standard injection method for mechanical feedback to inject the mass, energy and yields from stellar winds, and SNe \citep[][also described in appendix D in \citealt{hopkins18}]{hopkins14,kimm14,martizzi15,rosdahl17,hopkins18b}. This method accurately accounts for the conversion of energy into momentum done by ejecta reaching the resolved radii, and for the unresolved cooling. 

\subsection{Combined feedback output}

\begin{figure}
	\includegraphics[width=\hsize]{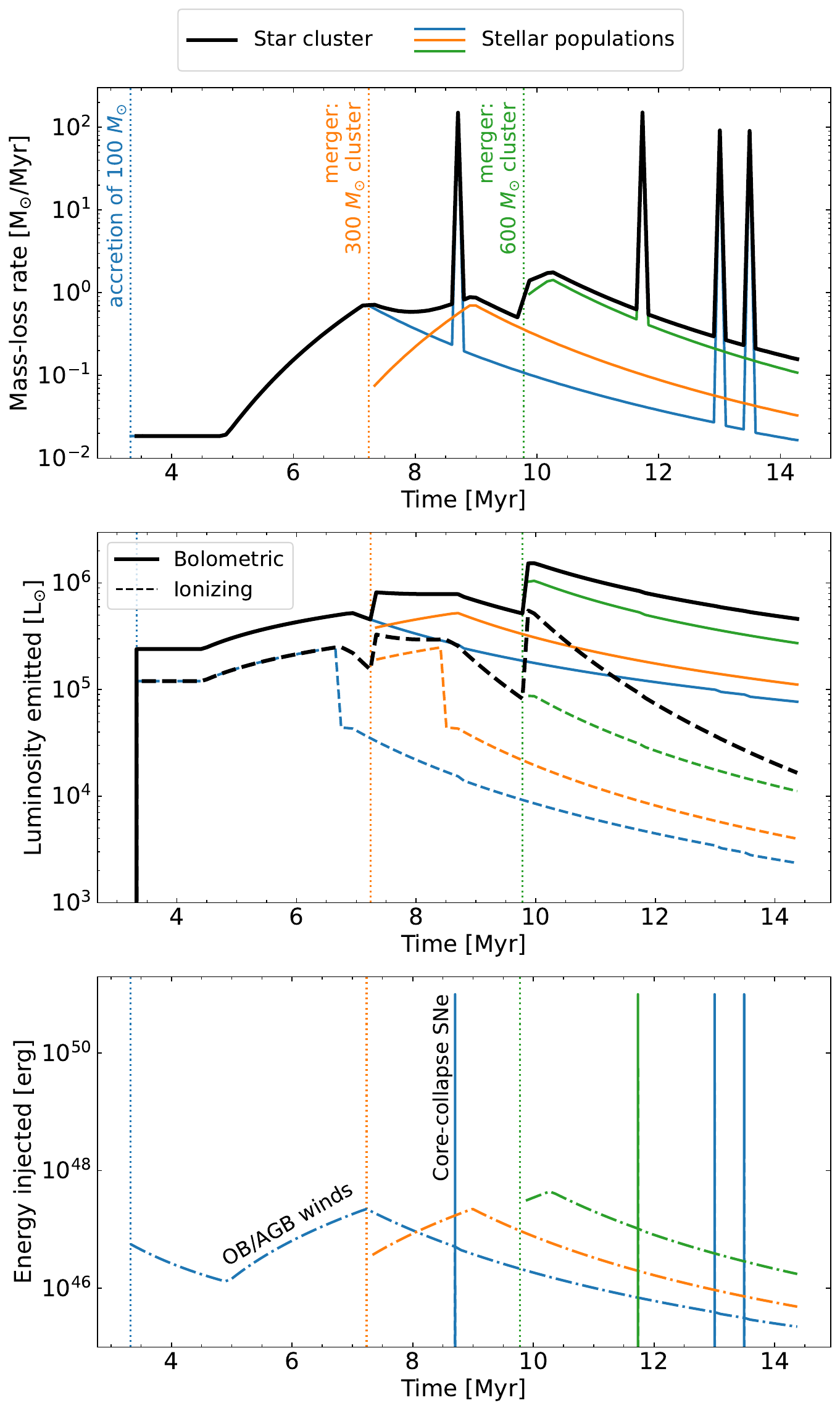}
  \caption{Evolution of the feedback products ejected by a representative star cluster (black line), with contributions of its three stellar populations shown as blue, orange, and green lines. \textit{Top:} mass-loss rate due to CC SNe and OB/AGB winds; \textit{Middle:} bolometric (solid) and ionizing (dashed lines) luminosities; \textit{Bottom:} energy injected by CC SNe events (solid lines) as well as the continuous kinetic energy injected due to OB/AGB mass loss (dash-dotted lines). Dotted vertical lines indicate events of mass growth for the star cluster and are labelled in the top panel.}
  \label{fig:vstime-fb-onecluster}
\end{figure}

Every star cluster $i$ is thus represented by a sink particle that can contain $j=1-20$ independent stellar populations. Each of the stellar populations within the star clusters is defined by an initial and current mass, age, total metallicity and yields. These quantities determine their evolution and the amount of feedback that each produces: the number of core-collapse and thermal SNe $N_{\rm CC}^{ij}$ and $N_{\rm Ia}^{ij}$, respectively, the bolometric luminosity $L_{\rm bol}^{ij}$, the masses ejected by supernovae and stellar winds, and the velocity of the wind mass loss. The combined feedback energy emitted by star clusters is thus the sum over the ejecta from its stellar populations,
\begin{equation}
\begin{aligned}
    E_{\rm CC}^{i} &= \sum_{j=0}^{19} N_{\rm CC}^{ij} E_{\rm CC}^{ij}\\ 
    E_{\rm Ia}^{i} &= E_{\rm Ia} \sum_{j=0}^{19} N_{\rm Ia}^{ij}\\
    E_{\rm w}^{i} &= \dfrac{1}{2} \sum_{j=0}^{19} M_{\rm w}^{ij} (v_{\rm w, inj}^{ij})^{2},
\end{aligned}
\end{equation}
where the number of core-collapse and thermal SNe and their energies, the mass ejected by the winds, $M_{\rm w}$, and its velocity, $v_{\rm w, inj}$, are defined in App.~\ref{app:fb}.
The total mass ejected is the sum over the ejected masses,
\begin{equation}
M_{\rm tot}^{i} = \sum_{j=0}^{19} (M_{\rm CC, ej}^{ij} + M_{\rm Ia}^{ij} + M_{\rm w}^{ij}),
\end{equation}
and similarly, the total mass in a species $k$ ejected is,
\begin{equation}
M_{{\rm tot}, k}^{i} = \sum_{j=0}^{19} (y_{{\rm CC}, k}^{ij}M_{\rm CC, ej}^{ij} + y_{{\rm Ia}, k}^{ij}M_{\rm Ia}^{ij} + y_{{\rm w}, k}^{ij}M_{\rm w}^{ij}),
\end{equation}
and the total velocity at which the mechanical feedback is injected is
\begin{equation}
    v_{\rm tot}^{i} = \sqrt{\dfrac{2}{M_{\rm tot}^{i}}\left[E_{\rm CC}^{i} + E_{\rm Ia}^{i} + E_{\rm w}^{i}\right]}. 
\end{equation}
Similarly, the radiative feedback emitted by the star cluster is the result of summing over the emission from the stellar populations within each band,
\begin{equation}
    L_{\nu}^{i} = \sum_{j=0}^{19} L^{ij} f_{\nu}^{ij}(t),
\end{equation}
where $f_{\nu}^{ij}(t)$ is the bolometric correction within a particular frequency range for the star population $j$ in the cluster $i$, which depends on the age of the stellar population.  

An example of the feedback emitted by a star cluster and its stellar populations is shown in Fig.~\ref{fig:vstime-fb-onecluster}. An initial mass growth shortly after the formation of the sink as a dense clump ($t\sim 3~\myr$, vertical dotted blue line in the top panel) leads to the formation of a single stellar population (blue line). This initial population is responsible for all the luminosity, mass and energy injected until $t\sim 7~\myr$, when a second stellar population formed at $t\sim5~\myr$ is added from a merged sub-cluster (orange line), and then at $t\sim 10~\myr$ a third population formed at $t\sim7~\myr$ is also added from a merged sub-cluster (green line). In this case, the metallicities of the stellar populations are identical, but their age differences triggers our prescription to keep each population separate within the star cluster. Increases to the stellar component of the cluster lead to corresponding increases in the total bolometric and ionizing luminosity that is being emitted (solid and dashed lines in the middle panel). The continuous mass loss from OB and AGB winds is also reflected in a continuous injection of kinetic energy of $10^{46}$--$10^{48}~\erg$, whereas each core-collapse SNe injects $10^{51}~\erg$ of energy instantaneously.

\section{Initial Conditions and Tests}\label{sec:ics-tests}

To test our numerical prescription, we evolve a suite of isolated spherical turbulent gas clouds using different assumptions, which we describe below. 

\subsection{Initial conditions}

We generate isolated gas clouds with the software \code{MakeCloud}\footnote{\href{https://github.com/mikegrudic/MakeCloud}{https://github.com/mikegrudic/MakeCloud}} using the `Sphere' mode. The resulting clouds are spheres of uniform density embedded in a diffuse gas with a density contrast of 1:1000, and we fix their mass to be $M_{\rm GMC} = 10^7~\msun$. Their radii are chosen to correspond to different initial gas surface densities, $\Sigma_{\rm GMC}$ (see Table~\ref{tab:sum-ics}). For our fiducial cloud, we use the molecular gas-weighted average gas surface density of molecular clouds in galaxy centers from the PHANGS collaboration \citep{sun22} of $\Sigma_{\rm GMC} = 210~\msunpc$. This density corresponds to a size of $R_{\rm GMC} = 123~\pc$. The other two clouds have increasingly smaller radii, $R_{\rm GMC} = 72.8, 51.5~\pc$, and correspondingly larger initial gas surface densities, $\Sigma_{\rm GMC} = 600, 1200~\msunpc$, respectively. 

The clouds are placed at the center of a periodic box of length $10\, R_{\rm GMC}$, and the gravity is kept non-periodic. We use an adaptive gravitational softening for the gas particles that is tied to their kernel length, with a minimum value of $\epsilon_{\rm gas} = 10^{-4} R_{\rm GMC}$, whereas sink particles have a fixed gravitational softening of $\epsilon_{\rm sink} = 1~\pc$. 

The initial velocity of the gas particles is drawn from a Gaussian random field with power spectrum $E_{k}\propto k^{-2}$ \citep{ostriker01}, and it is scaled to the turbulent virial parameter $\alpha_{\rm turb} = E_{\rm turb}/E_{\rm grav} = 0.9$. There is no driving maintaining the turbulence. The clouds are also assumed to be initially rotating with a virial parameter of $\alpha_{\rm rot} = E_{\rm rot}/E_{\rm grav} = 0.1$, such that the total virial parameter is $\alpha_{\rm vir} = \alpha_{\rm turb} + \alpha_{\rm rot} = 1$ and the clouds are bound. 

In order to resemble the natal conditions of massive clusters in the early Universe, we set the initial metallicity of the gas to be $\feh = -2$. We also set the initial gas and dust temperatures to be $20~$K, so they are initially in thermal equilibrium. The clouds are evolved with the $z=0$ UV background from \citet{faucher-giguere20}. 

\begin{table}
  \caption{Physical properties describing the turbulent clouds used as initial conditions in this work. From top to bottom, the table lists the mass, radius, initial gas surface density, virial parameter, velocity dispersion, initial free-fall timescale of the cloud, and minimum gravitational softening of the gas particles. }
  \label{tab:sum-ics}
  \centering{
    \begin{tabular}{l|ccc}
    \hline \hline
    Clouds & Fiducial & High $\Sigma$ & Highest $\Sigma$ \\ \hline 
    $M_{\rm GMC}$ [$\msun$] & $10^7$ & $10^7$ & $10^7$\\
    $R_{\rm GMC}$ [$\pc$] & 123 & 72.8 & 51.5\\
    $\Sigma_{\rm GMC}$ [$\msun~\pc^{-2}$] & 210 & 600 & 1200\\
    $\alpha_{\rm vir}$ & 1.00 & 1.00 & 1.00 \\
    $\sigma_{\rm GMC}$ [km/s] & 8.36 & 10.87 & 12.93\\
    $\tau_{\rm ff}$ [$\myr$] & 7.15 & 3.25 & 1.94  \\
    $\epsilon_{\rm gas}$ [$\pc$] & $1.23\times10^{-2}$ & $7.28\times10^{-3}$ & $5.15\times10^{-3}$ \\ 
    \hline \hline
    \end{tabular}}
\end{table}

\subsection{Tests}

\begin{table}
\centering{
\caption{Numerical experiments shown in this work. The bold font indicates the fiducial values. } \label{tab:sum-tests}
\begin{tabular}{ll}
 Parameter & Options tested \\  \hline \hline 
Growth of the clusters & \textbf{accretion and merging} \\  
 & only accretion \\  
 & only merging\\  
 & standard star formation\\  
Feedback mechanism & \textbf{SNe, winds and radiation}\\  
 & SNe and radiation\\  
 & SNe and winds\\  
 & winds and radiation\\  
 & no radiative pressure \\ 
 & no feedback\\  
 Initial gas surface & \multirow{2}*{$\Sigma_{\rm GMC} = \mathbf{210}, 600, 1200~\msun/\pc^2$}\\ 
density & \\ \hline
Mass resolution & \multirow{2}*{$m_{\rm res} = 10, \mathbf{100}, 1000, 10^{4}~\msun$}\\  
of the gas cells & \\ 
Rate of sink formation & $\mathbf{\epsilon_{\rm ff} = 100\%}$ \\ 
 & scaled the H$_2$ fraction\\
 & $\epsilon_{\rm ff} = 50\%$\\
 & $\epsilon_{\rm ff}(\alpha_{\rm vir}, \mathcal{M})$ \\
Gravitational softening & \multirow{2}*{$\epsilon = \mathbf{1~\pc}, 5~\pc$, and adaptive}\\ 
of the sinks & \\ 
\end{tabular}}
\end{table}

We explore the impact of different assumptions on the formation of star clusters in our prescription. In particular, we change the physical behaviour of star clusters, their feedback mechanisms, and the gas surface density of the cloud, as well as other numerical tests. In total, we evolve 19 clouds, which are summarized in Table~\ref{tab:sum-tests}. Unless otherwise stated, all the tests are performed for the cloud with an initial surface density of $\Sigma_{\rm GMC} = 210~\msunpc$ and mass resolution of $m_{\rm gas} = 100~\msun$.

In our first set of tests, we modify the mechanisms by which star clusters can grow mass over time. We consider cases in which sink particles can accrete gas and merge, either accrete or merge, and neither of them. In the latter case, sink particles thus correspond to single stellar populations of equal mass, as it is currently common in isolated and cosmological zoom-in simulations of Milky Way-mass galaxies and above. Therefore we call this test the `Standard star formation' case. These tests allow us to evaluate which mechanism, whether gas accretion or hierarchical merging of sub-clusters, is critical to regulate the growth of star clusters.

The second set of tests explore the impact of different combinations of feedback mechanisms in regulating the formation of star clusters. A large body of literature explores this topic \citep[e.g.][]{kim15,kim18,grudic18b,grudic21,li19b}. Here it is relevant to assess which mechanisms have a bigger impact on the growth of star clusters in our prescription, and what is their associated computing cost.

For our third set of tests we increase the initial gas surface density of the cloud, keeping the mass resolution fixed. 

\begin{figure*}
	\includegraphics[width=\hsize]{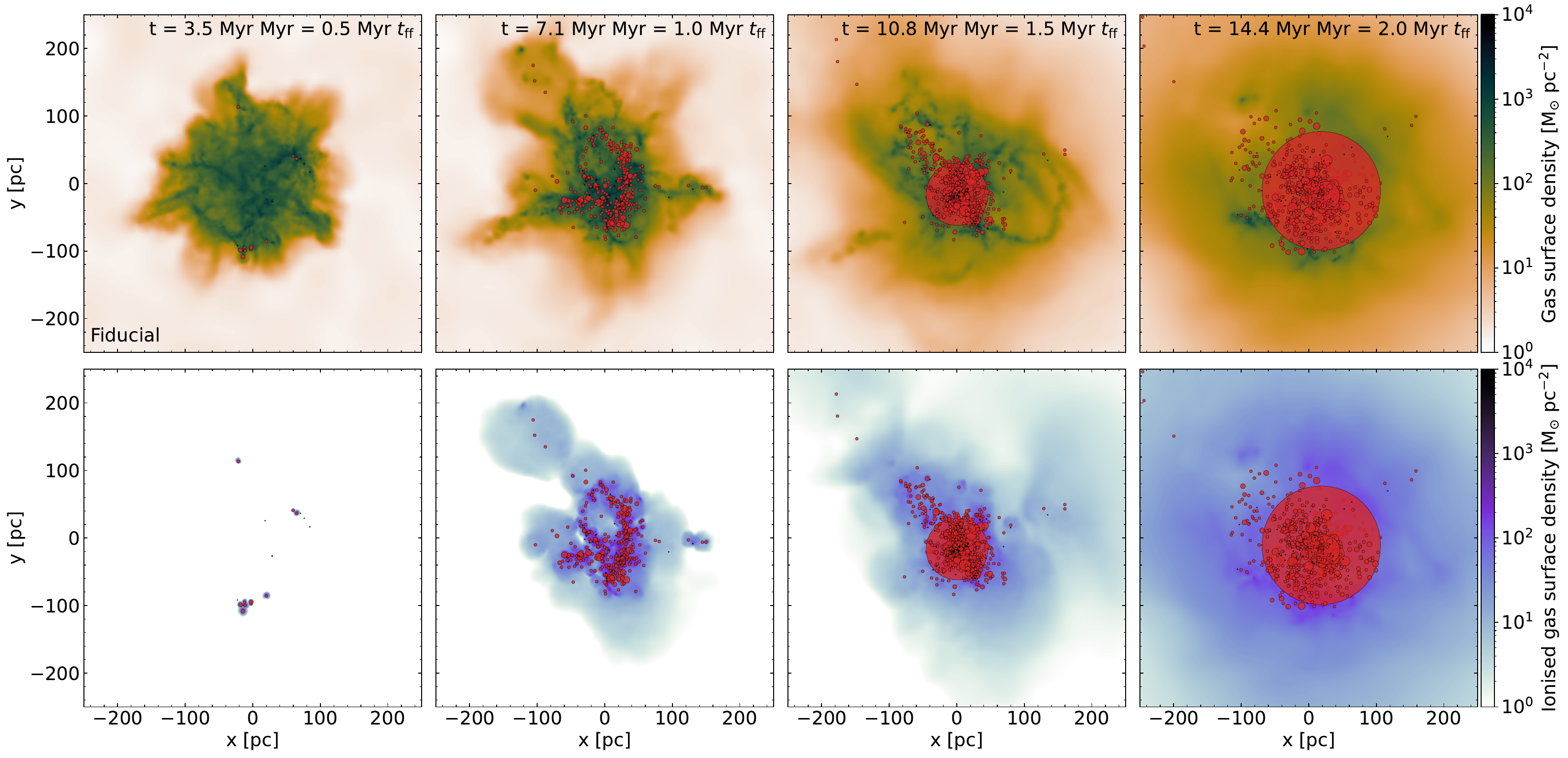}
  \caption{Time evolution of the projected gas surface density (\textit{top row}) and the projected ionised gas surface density (\textit{bottom row}) of the fiducial cloud at three representative times ($0.5t_{\rm ff}$, $t_{\rm ff}$ and $1.5t_{\rm ff}$). The star clusters are represented by red markers, scaled by mass. The black markers indicate sinks without stellar populations, i.e.~representing dense clumps of gas.}
  \label{fig:spatproj-fiducial}
\end{figure*}

Additionally, we explore the impact of purely numerical choices. Firstly, we change the mass resolution of the cloud to $m_{\rm gas} = 10, 10^3, 10^4~\msun$. In these tests, we vary the density threshold for sink formation $n_{\rm th}$ accordingly by the same amount. Secondly, we modify the rate at which sink particles form, $\dot{m}_{\star}$. In one case, we include an additional criterion for sink formation in which the star formation rate is scaled by the fraction of molecular gas in the gas particle, $\dot{m}_{\star} = f_{\rm H_2}\epsilon_{\rm ff} m_{\rm gas}/t_{\rm ff}$. The two other cases modify the assumed star formation efficiency per free-fall time, $\epsilon_{\rm ff}$: in one of them it is lowered to a constant $50~$per cent, whereas in the other the turbulence of gas is allowed to regulate star formation (based on \citealt{federrath12,federrath13,hopkins13}, and similar to \citealt{kretschmer20}). Thirdly, we change the gravitational softening of the sinks, $\epsilon_{\rm sink}$, as this parameter affects the rate of sub-cluster merging by defining the region within which clusters are considered to be bound. The evolution of the clouds under these final assumptions, as well as the resulting star clusters, are very similar to the fiducial cloud. Thus, to clarify the discussion we only show them in Appendix~\ref{app:numerical-tests}. 

\section{Results}\label{sec:results}

We describe first the evolution of the fiducial cloud, as well as how different assumptions impact the formation efficiency and the cluster mass function. Then, we explore whether our clustered feedback prescription significantly alters the evolution of clouds compared to the standard assumption of star formation. For that, we focus on the effect of resolution, and on the properties of the gas from which star clusters form, and where they deposit their feedback energy.

\subsection{Evolution of the fiducial cloud}

\begin{figure}
    \includegraphics[width=\hsize]{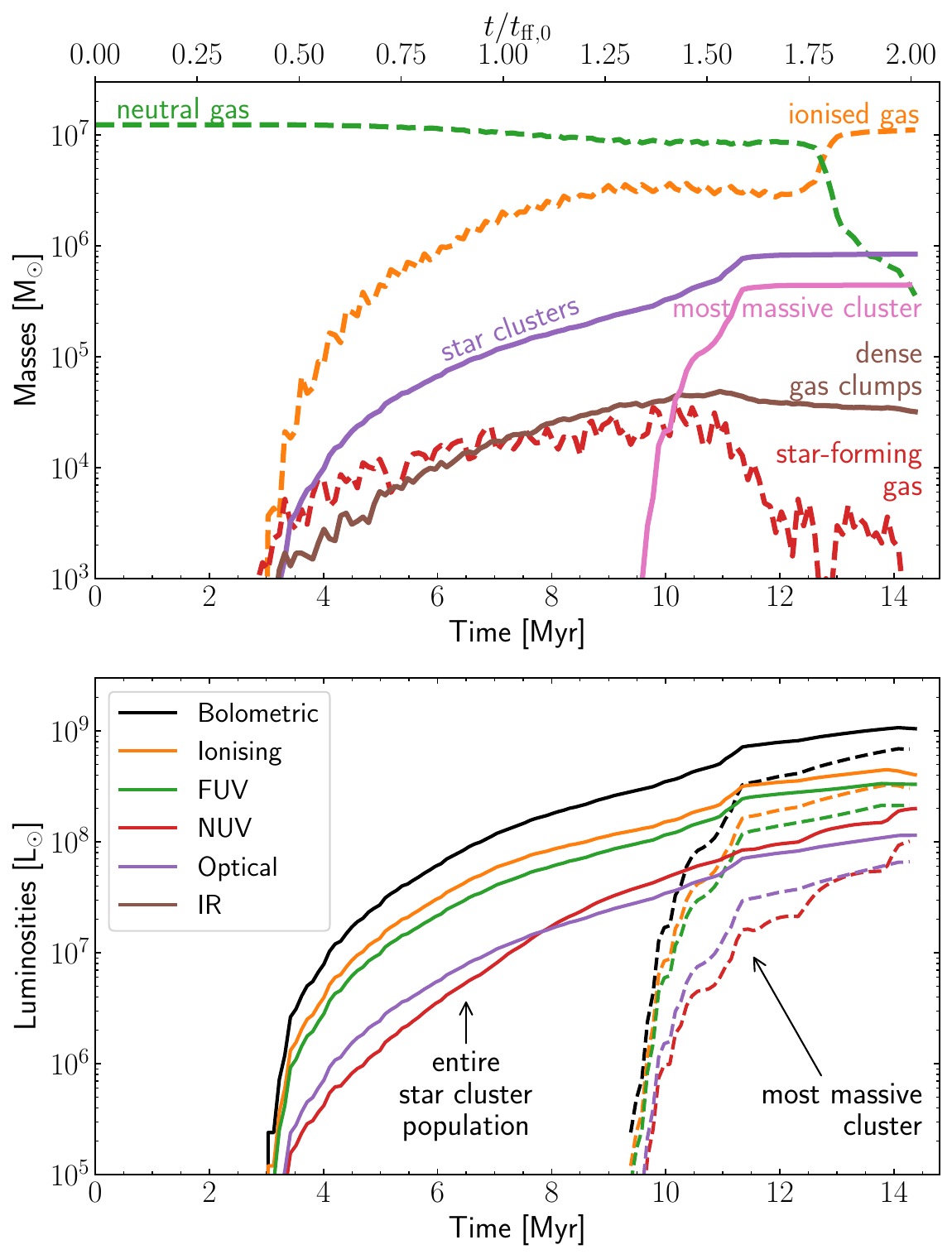}
  \caption{Evolution of the fiducial cloud: mass (\textit{top panel}), and luminosity (\textit{bottom panel}). Lines are labelled within the panels, and we emphasize the contribution of the most massive cluster in this cloud.}
  \label{fig:vstime-fiducial}
\end{figure}

Upon starting the simulation, the fiducial cloud collapses under self-gravity developing a filamentary structure in which denser gas cells are located in the filaments and their nodes. The first sink is created as a dense clump of gas at $t=2.84~\myr$, and a second forms at $t=2.88~\myr$. Both of them accrete gas particles, and shortly after at $t=2.95~\myr$, the first star cluster forms within the second sink. After $0.5t_{\rm ff}$ of evolution, there are ten star clusters with masses ranging between $300$--$600~\msun$, with a total of $3900~\msun$. The clusters are located predominantly in two high-density regions, located towards the bottom and top-right of the cloud. Additionally, there are $1700~\msun$ in starless sinks that have not yet formed stellar populations (top-left panel in Fig.~\ref{fig:spatproj-fiducial}).  

As the cloud continues to collapse, the star clusters grow in mass by accreting gas and merging with less massive ones (top panel in Fig.~\ref{fig:vstime-fiducial}). After a free-fall time of evolution, there are $300$ star clusters with masses between $300$--$2400~\msun$, and they contain a total of $1.9\times10^5~\msun$. The majority of them are located along a horse-shoe shape following the gas distribution of the cloud (second column in Fig.~\ref{fig:spatproj-fiducial}), with a few seemingly being ejected from the cloud towards the top left corner of the domain. In addition, there are $1.74\times10^4~\msun$ of dense gas clumps without stellar populations. The number of star clusters rises to $650$ with masses $300$--$1.2\times 10^5~\msun$ at $t=1.5t_{\rm ff}$, and slightly decreases to $621$ star clusters with masses $300$--$4.45\times 10^5~\msun$ at the end of the simulation ($t=2t_{\rm ff}$, right-hand column in Fig.~\ref{fig:spatproj-fiducial}). Out of the initial gas mass of $10^7~\msun$, at the end there is a total of $8.44\times10^5~\msun$ in star clusters, and $3.2\times10^4~\msun$ of dense gas clumps.

Shortly after their formation, the first stellar populations start radiating ionising luminosity and injecting continuous OB wind mass and momentum into the gas (bottom panel in Fig.~\ref{fig:vstime-fiducial}). Initially, the bubbles of ionised gas are confined around the star clusters (bottom-left panel in Fig.~\ref{fig:spatproj-fiducial}), but as the amount of ionising radiation increases from subsequent star clusters, it grows into roughly the initial size of the cloud by one free-fall timescale and it dominates the domain at the end of the simulation (bottom-right panel in Fig.~\ref{fig:spatproj-fiducial}).

Despite having only an initial mass of $10^7~\msun$, several of our clouds form very massive star clusters. In the case of the fiducial cloud, at the end of the simulation a massive star cluster of $4.45\times 10^5~\msun$ has formed, which corresponds to $4.45\%$ of the initial cloud mass. The massive cluster in our fiducial cloud contains 9 different stellar populations, some of which form in-situ after accreting gas and some that have been retained from less massive clusters which merged with the larger cluster. The luminosity emitted by its stellar populations quickly dominates the luminosity radiated by the entire star cluster population (bottom panel in Fig.~\ref{fig:vstime-fiducial}), and it significantly reduces the amount of available  star-forming gas by heating it up. The cluster forms relatively late in the evolution of the cloud, at $t=9.3~\myr=1.3t_{\rm ff}$, and within $\sim0.2~\myr$ it has gained $\sim2\times10^3~\msun$ from merging with less massive clusters \citep[see also][]{howard18,dobbs22}. This rapid mass growth increases its predicted gas accretion rate, which peaks at $1.2~\msun/\yr$ roughly $1.8~\myr$ after formation. 

\subsection{Formation efficiencies}

\begin{figure*}
  \includegraphics[width=\hsize]{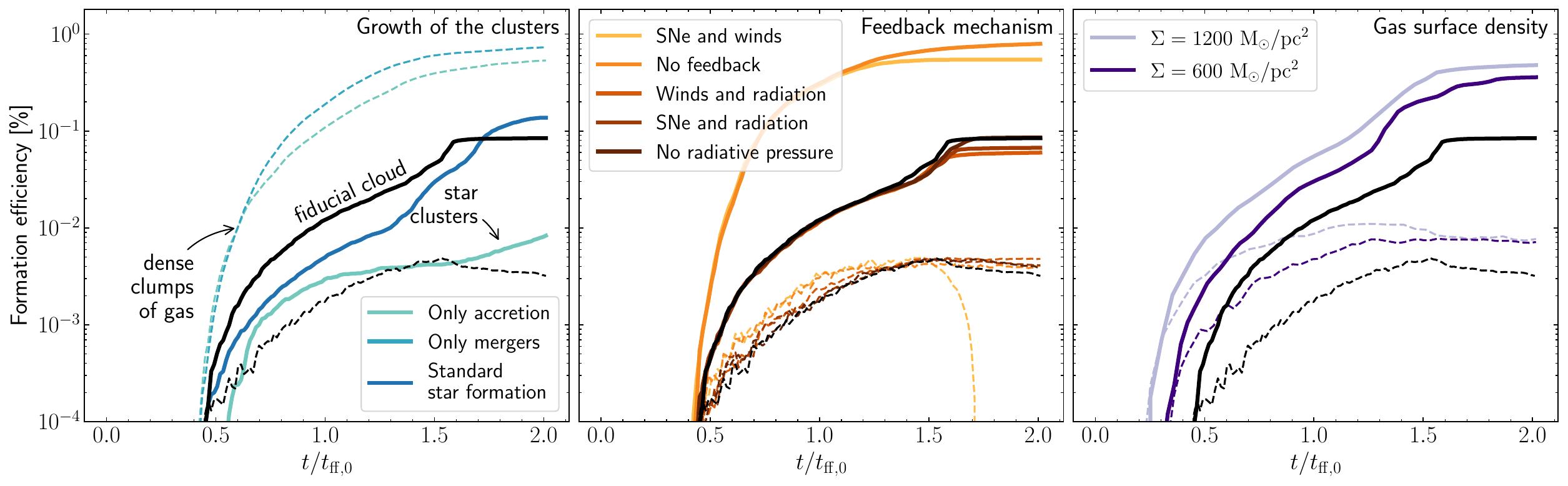}
  \caption{Formation efficiency of star clusters (thick solid line) and of sinks that remain starless (dashed thin line) amongst our suite of clouds as a function of the initial free-fall timescale of the cloud. From left to right, we show clouds in which we modify the growth of star clusters, their feedback mechanisms, as well as the initial gas surface density of the cloud. The black solid line present in all panels corresponds to our fiducial cloud.}
  \label{fig:formEfficiency-allruns}
\end{figure*}

We explore here the efficiency of forming star clusters, which we show as a function of time in Fig.~\ref{fig:formEfficiency-allruns}. The total or integral star cluster formation efficiency in each of our clouds is measured at $t=2t_{\rm ff}$ as
\begin{equation}
\epsilon_{*} = \dfrac{M_*(t=2t_{\rm ff})}{M_{\rm GMC} (t=0)}.
\end{equation}

Through most of its evolution, the fiducial cloud forms mass in clusters more quickly than the cloud evolved with the standard star formation prescription, but cluster formation stops at $t\simeq12~\myr$ (left panel in Fig.~\ref{fig:formEfficiency-allruns}). In contrast, the cloud with standard star formation continues forming stars until the end of the simulation, and its formation efficiency ends up being almost a factor of two larger than in the fiducial cloud ($\epsilon_{*} = 0.14$ vs $\epsilon_{*} = 0.08$). The cloud evolved assuming that clusters can only grow by accreting gas experiences low formation efficiencies overall ($\epsilon_{*} < 0.01$). In a more extreme case, the cloud in which clusters can only grow from mergers with other sinks forms no star clusters, and $70\%$ of the initial cloud mass remains in dense clumps of gas. Thus, both gas accretion and hierarchical merging with sub-clusters are critical processes that drive the formation of star clusters.

Radiation, and in particular, ionising radiation, is the main feedback mechanism regulating the formation of star clusters in our clouds. UV radiation ionises and heats up the gas around the clusters, which increases the sound speed of the gas, and decreases the accretion rate onto the star clusters (eq.~\ref{eq:bondi-hoyle-mdot}). Over time, the hot ionised bubbles of gas expand, causing their density to decrease. Subsequent feedback deposited by the star clusters can act more efficiently in these lower density surroundings than if radiation had not acted. Neglecting the effect of radiative pressure leads to a similar evolution, indicating that the bubbles expand from their increased temperature, and not from the momentum exchange with the photon field. In the presence of radiative feedback, the onset of SNe does not significantly alter the formation efficiency of the star clusters. The number of starless sinks is similar regardless of which feedback mechanisms are active. In our sample, mechanical feedback becomes inefficient due to the low metallicity of the clouds. Although we do not include that the more massive end of the IMF fails to explode as core-collapse SNe, the weaker stellar winds and the effect of catastrophic radiative cooling lead to radiative feedback dominating the formation of star clusters. 

Lastly, the clouds with initially higher gas surface density form increasingly more mass in star clusters. This increasing trend with gas surface density has been found in many previous studies where star particles are single stellar populations that do not accrete nor merge \citep[e.g.][]{grudic18b, kim18, li19b}, and it is due to feedback increasingly becoming more inefficient at regulating the formation of stars in higher density environments. From these comparisons, we find that both gas accretion and hierarchical merging, together with the emission of radiative feedback, are critical ingredients to model the formation of star clusters.

\subsection{Timescales of gas depletion and star cluster formation}

In the majority of our clouds, star cluster formation starts $\sim0.5t_{\rm ff}$ after the initial collapse of the clouds. 
To determine whether the gas is being exhausted by the formation of clusters, we estimate the gas depletion timescale as the ratio of the initial GMC mass to the average cluster formation rate over the evolution of the cloud:
\begin{equation}
t_{\rm dep} = \dfrac{M_{\rm GMC}}{<\rm SFR>}
\end{equation}
The depletion time can then be compared to the time over which feedback can clear out neutral gas to understand what regulates the overall star formation efficiency. As a proxy for the final state of the cloud, we use the fraction of neutral gas present in its vicinity, $r<r_{\rm GMC}$. 

Overall, we can divide our clouds in three final states: star cluster formation would consume all the gas (low depletion timescales), feedback from the star clusters has cleared out the cloud from neutral gas (long depletion timescales and low fractions of neutral gas left), and feedback has not acted yet on the evolution of the cloud (long depletion timescales and high fractions of neutral gas left). Among our sample of clouds, the depletion timescale ranges between $t_{\rm dep} \sim 2$--$110.2~t_{\rm ff}$, and the fraction of neutral hydrogen lies between $f_{\rm HI}\sim 0$--$0.25$.

The clouds without feedback or without ionising radiation and with higher gas surface densities would be able to consume the entirety of the initial cloud mass within $<5t_{\rm ff}$. In contrast, the clouds that include ionising radiation are capable of reducing the amount of neutral gas to $<10~$per cent. The only cloud whose evolution is not affected yet by feedback is the one evolved with the standard representation of star formation, which retains the largest fraction of neutral gas, $f_{\rm HI} = 0.25$. The inclusion of ionising radiation is thus the critical factor setting the fate of the clouds.

A unique feature of our numerical prescription is that we keep track of the different stellar populations within each star cluster, which allows us to quantify their internal age spreads. Compared to previous studies in which star clusters are described with a single age \citep{li17,howard18,brown22}, we can estimate the duration of star cluster formation and provide constraints to theoretical models of chemical enrichment in star clusters \citep[e.g.][]{cottrell81,denissenkov14,bastian18,nguyen24}.

\begin{figure}
  \includegraphics[width=\hsize]{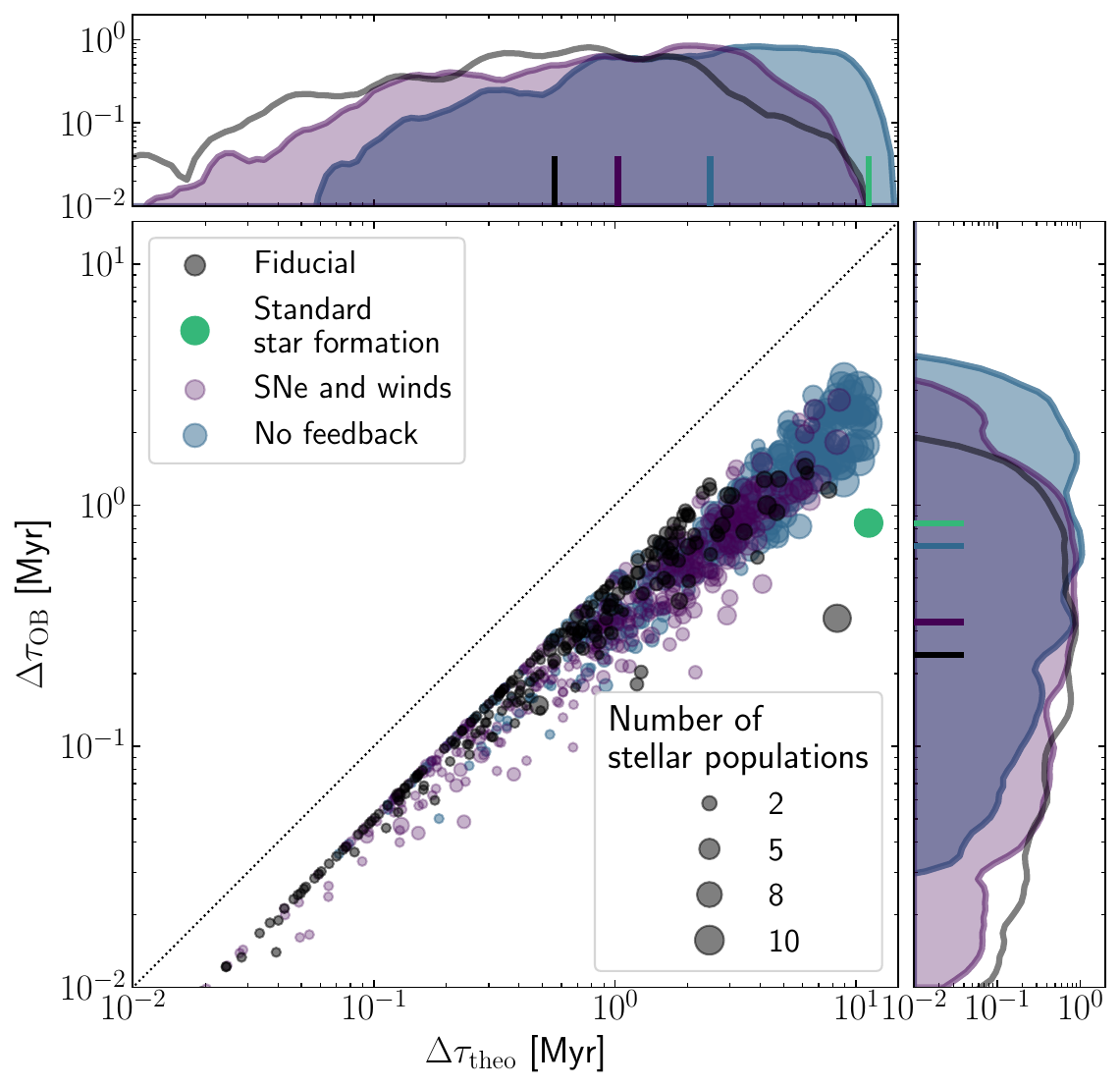}
  \caption{Observational age spread as a function of the theoretical spread for all star clusters in four representatives clouds. The size of the markers corresponds to the mass of the star clusters, and the black dotted line marks the 1:1 relation. Panels on the sides show the KDEs, with the horizontal and vertical solid lines indicating the median of each distribution.}
  \label{fig:delta-tau-theoretical-ionlum}
\end{figure}

To quantify the internal age spreads, we use the mass-averaged ages, masses and metallicities of every formation event within the star clusters. These formation events can be triggered either by the accretion of gas or by merging, and they reflect the stars that end up composing the cluster. We estimate two different internal age spreads of the star clusters. Firstly, we consider a theoretical age spread,
\begin{equation}
\Delta \tau_{\rm theo} = \tau_{\rm oldest} - \tau_{\rm youngest},
\end{equation}
defined as the age difference between the youngest and oldest formation events. This definition describes the total duration of star cluster growth, and it is relevant to understand how long internal processes such as chemical enrichment can occur. Secondly, we determine the standard deviation of the ages of the formation events weighted by their ionizing luminosity,
\begin{equation}
    \Delta \tau_{\rm OB}^2 = \dfrac{1}{\sum_j L_{j, \rm ion.}} \sum_j \left[ L_{j, \rm ion.} (\tau_j - <\tau>_{L_{j, \rm ion.}})^2 \right],
\end{equation}
where $<\tau>_{L_{j, \rm ion.}} = \left( \sum_j L_{j, \rm ion.} \tau_j\right)/\sum_j L_{j, \rm ion.}$ is the ionising luminosity-weighted mean age. This definition mimics the observational age spreads measured from massive stars in young star clusters \citep[e.g.][]{martocchia18b}. Since a substantial fraction of the star clusters have a single burst of star formation, we only show in Fig.~\ref{fig:delta-tau-theoretical-ionlum} the age spreads for clusters with multiple formation events. 

The median observational age spreads of the star clusters among our sample of clouds are $0.1$--$0.8~\myr$, consistent with observational constraints \citep[e.g.][]{cabrera-ziri14,martocchia18b}. The median theoretical age spreads are a factor of $2$--$3$ larger than the observational spreads, ranging between $0.3$--$2.6~\myr$. Two sets of clouds stand out from our sample. In the first, the cloud evolved with the standard star formation prescription forms a single bound cluster\footnote{We identify star clusters as self-gravitating structures with the implementation of DBSCAN \citep{ester96} within the python \code{sklearn} library. Clusters are required to have at least six members, a maximum distance between core points of $1~\pc$ and be gravitationally bound based on their total energy. We find a single star cluster of $m=1.32\times10^6~\msun$ present after $2~t_{\rm ff}$ in the cloud evolved with standard star formation, surrounded only by unbound stars.}, and its age spreads agree with the longest spreads in the fiducial cloud. The second set corresponds to the clouds evolved without radiative feedback, where the lack of ionising radiation leads to clusters forming over longer timescales \citep[see][]{rathjen21,andersson24,deng24}. 

\subsection{Cluster mass functions}

\begin{figure*}
	\includegraphics[width=\hsize]{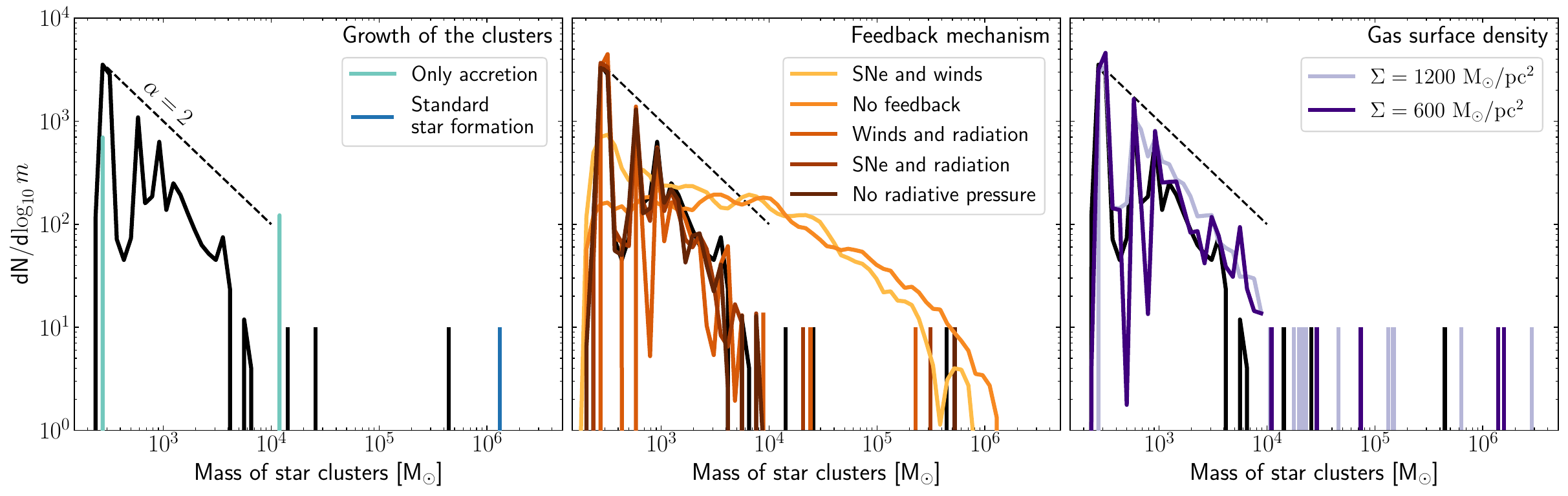}
  \caption{Mass function of star clusters in our entire suite of clouds calculated as a KDE with an Epanechnikov kernel. From left to right, we show clouds in which we modify the growth of star clusters, their feedback mechanisms, as well as the initial gas surface density of the cloud. The black solid line present in all the panels corresponds to our fiducial cloud, and the thin dashed lines provide a visual representation of ${\rm d}N/{\rm d}m \propto m^{-2}$ between $300$--$10^4~\msun$. Star clusters more massive than $10^4~\msun$ are shown individually as vertical lines.}
  \label{fig:dndlog10m-allruns}
\end{figure*}

We explore here the masses of the star clusters formed among our sample of clouds at the end of the simulations ($t=2t_{\rm ff}$, Fig.~\ref{fig:dndlog10m-allruns}). Given that our cluster populations form from a single cloud of $10^7~\msun$, the formation of clusters more massive than $10^4~\msun$ is highly stochastic. We estimate the mass functions of low-mass star clusters using a kernel density estimate (KDE) with the Epanechnikov kernel, but show clusters with $m>10^4~\msun$ as single vertical lines in Fig.~\ref{fig:dndlog10m-allruns}. We also characterise the shape of the mass functions by fitting a power-law to the mass range $300$--$10^4\,\msun$.

Observations of young star clusters find that their masses are well characterized by power-law functions, with some suggestion that the high-mass end might be exponentially cut \citep[e.g.][]{portegies-zwart10,krumholz19,adamo20}. These observational measurements, together with theoretical expectations of collapse in a turbulent medium, suggest that the slope of the power-law part of the mass function is ${\rm dN}/{\rm d}m \propto m^{-\alpha}$ with $\alpha = 2$ \citep[e.g.][and references above]{elmegreen11}.
We find that the slope of the cluster mass function in the fiducial cloud is $\alpha = 2.1\pm0.3$, close to the observational and theoretical expectations. Among our sample of clouds, the biggest change on the mass function is caused when radiative feedback is not included. As discussed in detail in \citet{andersson24} and \citet{deng24}, the presence of pre-supernova feedback limits the mass growth of star clusters, thus setting the steep power-law shape and reducing the cluster formation efficiency. Including the effect of ionising radiation is therefore critical to regulate the growth of star clusters.

\subsection{Clustered vs standard star formation and feedback}\label{sub:clustered-standard-sf}

\begin{figure*}
	\includegraphics[width=\hsize]{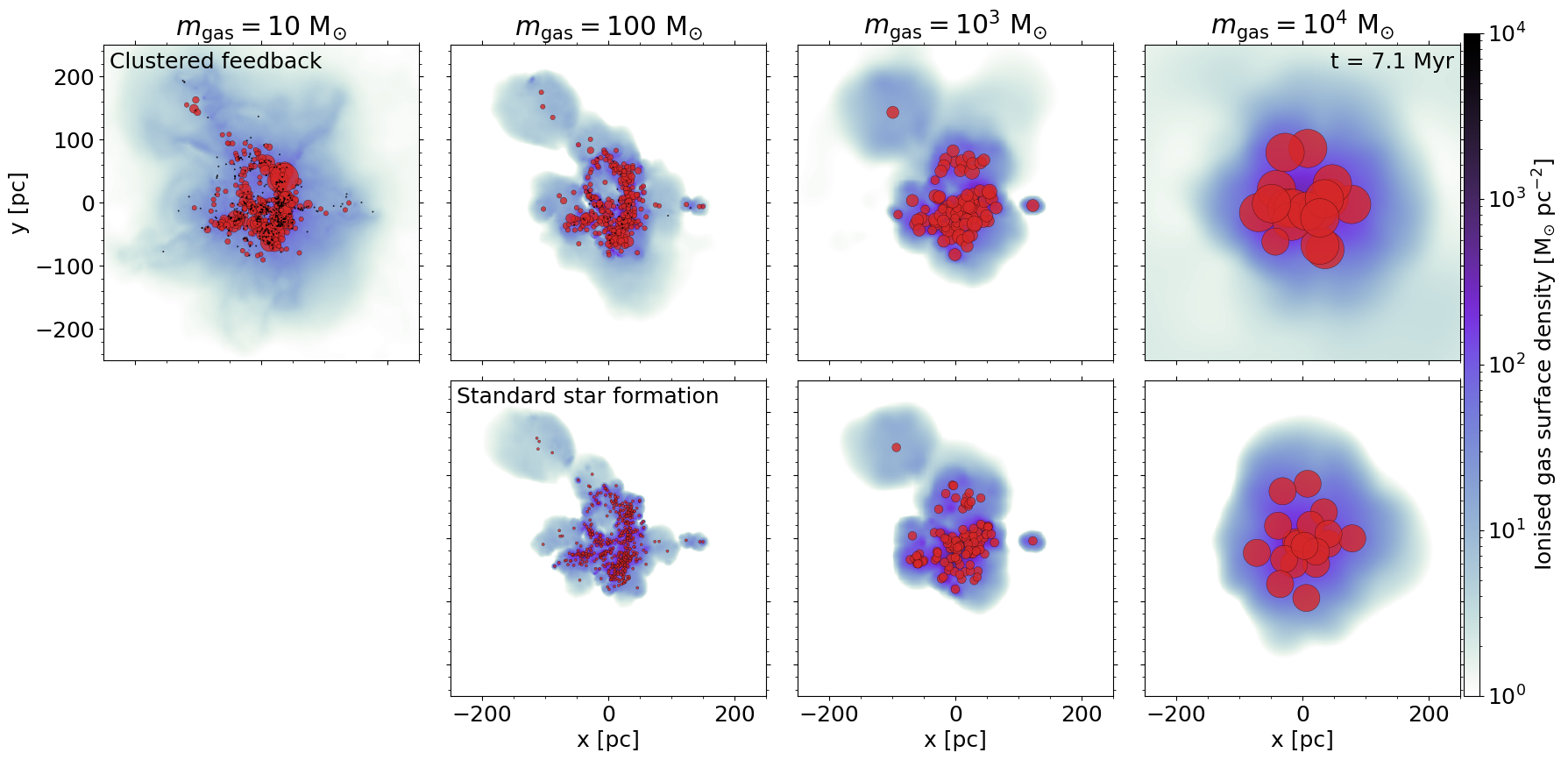}
  \caption{Projected ionised gas surface density distributions of the same turbulent cloud of $M_{\rm GMC} = 10^7~\msun$ at four mass resolutions (from left to right, $m_{\rm gas} = 10, 100, 10^3, 10^4~\msun$, respectively) evolved with our clustered feedback prescription (\textit{top row}) and with the standard formalism for star formation and feedback (\textit{bottom row}). The red markers represent the star clusters in the top, and star particles in the bottom row, respectively, and are scaled by their mass. Black markers in the top row indicate sinks without stellar populations, i.e.~representing dense clumps of gas. }
  \label{fig:spatproj-mgas-cluster-sf}
\end{figure*}

\begin{figure}
	\includegraphics[width=\hsize]{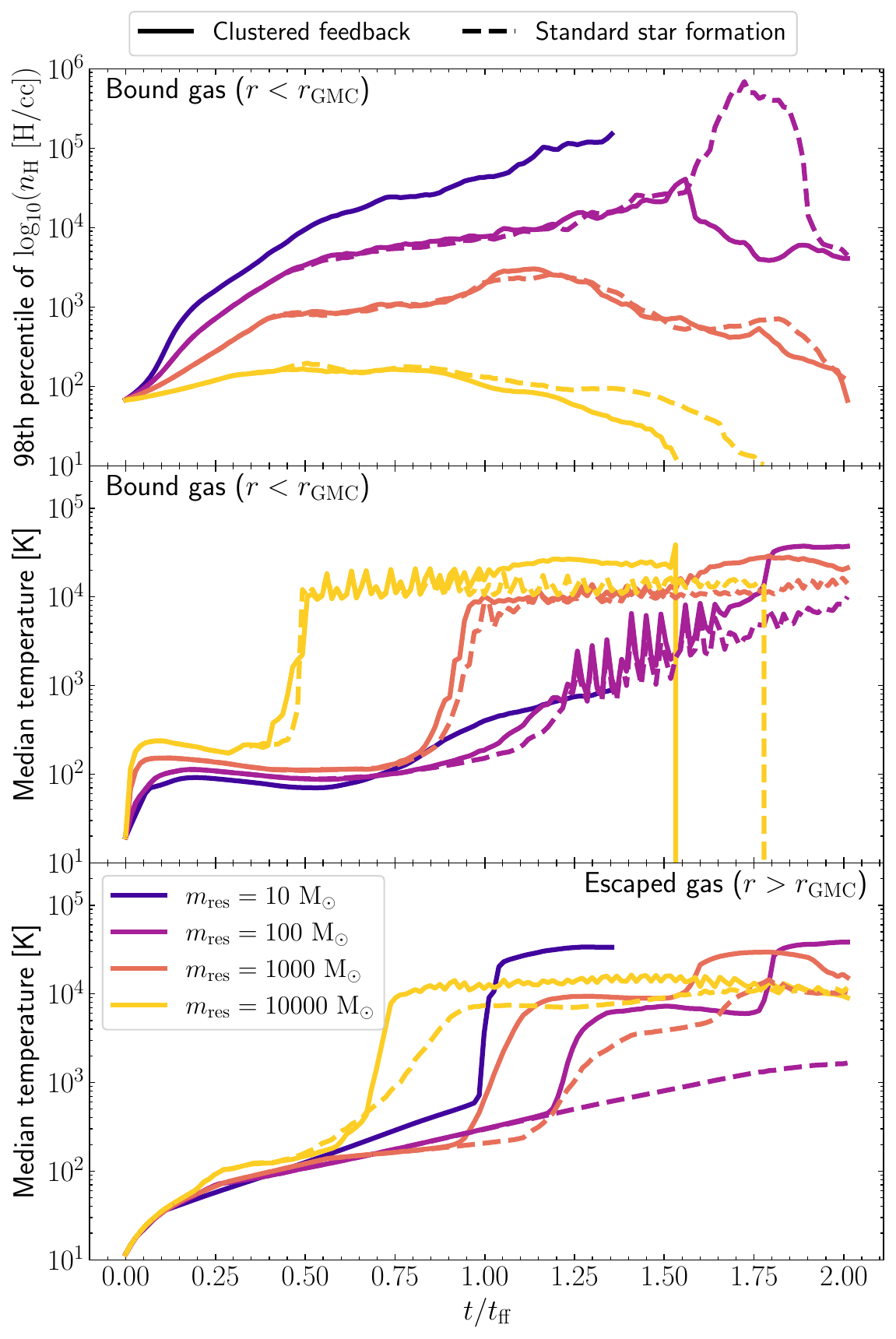}
  \caption{Comparison of the effect of the clustered feedback prescription (solid lines) relative to the standard model of star formation (dashed lines): panels show the time evolution of the $98$th percentile of the bound gas density distribution (\textit{top}), and the median temperatures among the bound ($r<r_{\rm GMC}$, \textit{middle}) and escaped gas ($r>r_{\rm GMC}$, \textit{bottom}). Line colours correspond to different gas mass resolution.}
  \label{fig:vstime-mgas-cluster-sf}
\end{figure}

\begin{figure}
	\includegraphics[width=\hsize]{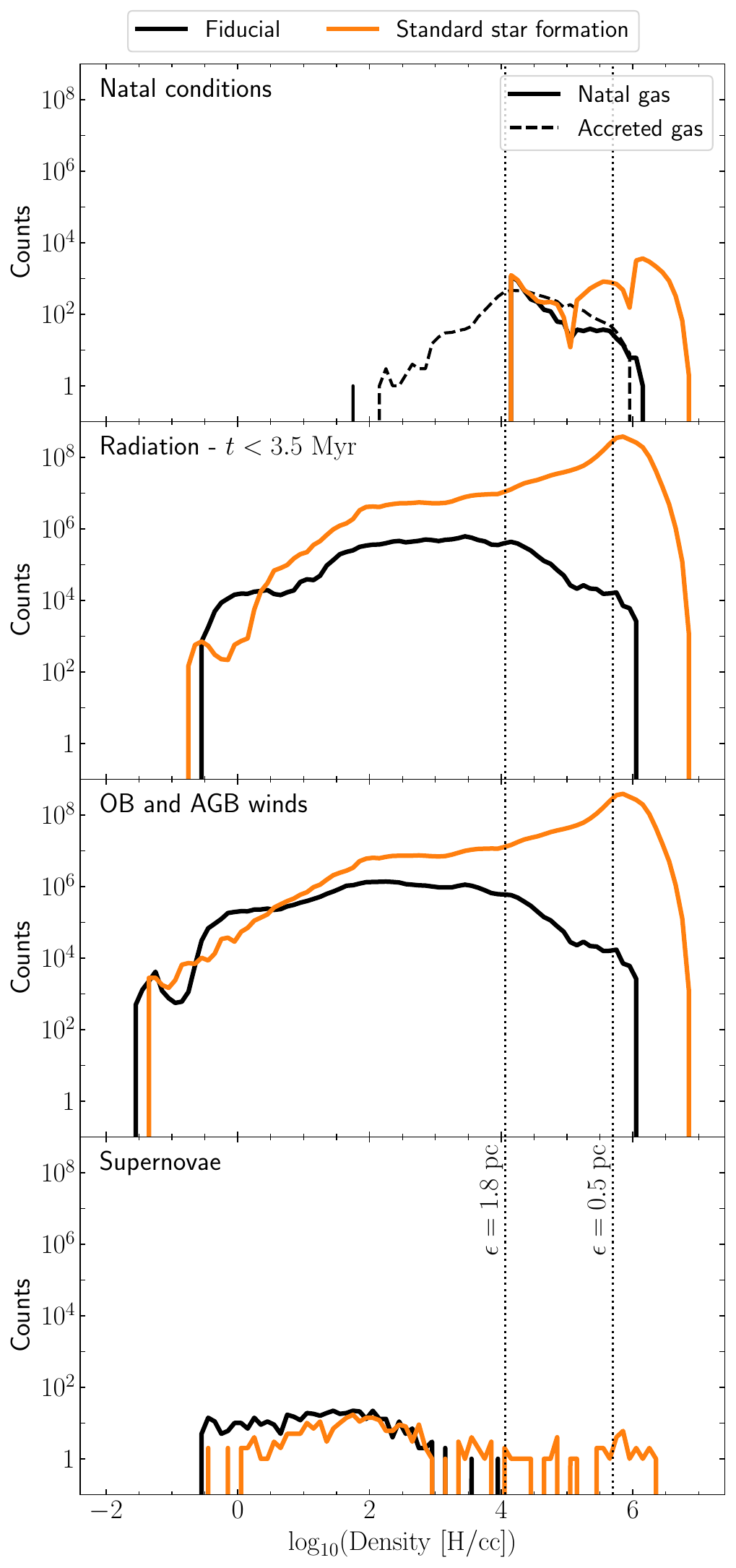}
  \caption{Gas density distributions: (\textit{top}) densities in which sinks form, as well as the density of the accreted gas, (\textit{second}) densities in which most of the ionizing radiation is emitted, (\textit{third}) densities in which stellar winds are ejected, and (\textit{bottom}) densities in which SNe explode. The vertical dotted lines mark the critical densities supported by the two gravitational softening lengths of the gas particles.}
  \label{fig:pdf-gas-rho-fb}
\end{figure}

Now that we have assessed the impact of different choices on the formation of star clusters, we explore whether our clustered feedback prescription significantly alters the evolution of the clouds. Since our model is an alternative approach to the standard assumption that star particles represent single stellar populations of equal mass (i.e.~the standard star formation), we use the `Standard star formation' cloud as a benchmark. 

The goal of our prescription is to model individual star clusters while bridging the scales between high-resolution single-star simulations of isolated clouds and dwarf galaxies and cosmological zoom-in simulations of galaxy formation. Thus, we focus on the effect of changing the mass resolution on the evolution of the clouds (Fig.~\ref{fig:spatproj-mgas-cluster-sf}). 

Overall, we find that the clouds undergo very similar evolution at fixed gas mass resolution. Their total formation efficiencies are $\epsilon_{*}\sim10~$per cent at $m=100~\msun$, and $\epsilon_{*}=3$--$4$ and $2$--$4$~per cent at $m=10^3~\msun$ and at $m=10^4~\msun$ for the clustered and standard star formation clouds, respectively. Regardless of the assumption on star formation, the clouds tend to evolve faster for increasingly lower mass resolution. At lower resolutions, the relevance of sub-cluster mergers decreases and clusters grow in mass primarily from gas accretion. 

Despite forming similar amounts of stellar mass, the clouds evolved with clustered feedback tend to heat up and disperse faster than when they are evolved with the standard formalism. At late times in the simulations ($t>1.25t_{\rm ff}$), the clouds evolved with the standard prescription tend to develop pockets of high-density gas (upper panel in Fig.~\ref{fig:vstime-mgas-cluster-sf}), whereas gas accretion in the clouds with clustered feedback transforms gas into stellar populations before it can collapse to higher density. In the clustered-feedback clouds, this leads to feedback being more efficient in dispersing the natal environments by heating up the gas within the cloud (middle panel in Fig.~\ref{fig:vstime-mgas-cluster-sf}), as well as outside the cloud (lower panel in Fig.~\ref{fig:vstime-mgas-cluster-sf}). This creates channels for the radiation to escape, and the clouds and their surrounding medium become quickly ionised. Already after a free-fall time of evolution, the projected ionised gas surface densities (Fig.~\ref{fig:spatproj-mgas-cluster-sf}) show that the ionising fronts in the clustered feedback clouds have reached further out than when using the standard star formation model.

\subsection{Sites of cluster formation vs sites of feedback release}

To understand why the clouds evolved with the clustered and standard formation assumptions develop pockets of gas of different densities, we examine the gas densities in which star clusters form, and in which different feedback mechanisms are released (Fig.~\ref{fig:pdf-gas-rho-fb}). At the end of the simulations ($t=2t_{\rm ff}$), the bound gas in the fiducial cloud has evolved into lower densities than the one in the standard star formation, with median gas densities of $38$ and $100~\hcc$, respectively. 

The fiducial cloud forms sinks in gas with $n_{\rm H}\sim10^4$--$10^6~\hcc$, and these sinks then accrete gas particles with densities in the range $n_{\rm H}\sim10^2$--$10^6~\hcc$. The process of gas accretion provides another pathway for converting gas particles into stellar populations, which removes particles that would otherwise collapse to higher densities. In this cloud, we find that feedback mechanisms that act on increasingly longer timescales inject their mass, energy, momentum and radiation in environments of decreasing gas density. Thus, the ionising radiation is injected in gas particles with densities ranging $n_{\rm H}\sim0.3$--$10^6~\hcc$, whereas the kinetic energy from OB winds is emitted in $n_{\rm H}\sim3\times10^{-2}$--$10^6~\hcc$ and the SNe deposit their energy in $n_{\rm H}\sim0.3$--$10^3~\hcc$. Pre-SNe feedback is therefore critical for clearing out the natal environment around the star cluster, which leads to more efficient SNe feedback \citep[e.g.][]{mcleod21, chevance22}. 

In contrast, the cloud evolved with the standard star formation prescription undergoes a period of intense and bursty star formation around $t\sim1.6$--$1.9t_{\rm ff}$. During this period, the $98$th percentile of the gas distribution increases from $n_{\rm H}=3\times 10^4~\hcc$ to $4\times 10^5~\hcc$ (see Fig.~\ref{fig:vstime-mgas-cluster-sf}), and a significant fraction of stars form during that period, with natal gas densities ranging $n_{\rm H}\sim10^{4}$--$10^7~\hcc$. This higher range of natal gas densities leads to the different feedback mechanisms being injected primarily into gas particles with densities $n_{\rm H}\sim10^6~\hcc$, causing them to be highly inefficient at dispersing the gas. By the time SNe explode, they deposit their energy and momentum in densities $n_{\rm H}\sim1$--$10^6~\hcc$, which is two orders of magnitude higher than in the case of the fiducial cloud. Thus, a major advantage of using a sink-based approach for modelling star clusters is that gas accretion into the clusters prevents the development of pockets of very high-density gas, which is key for a more numerically stable prescription.

\section{Discussion}\label{sec:discussion}

We briefly compare our total formation efficiencies with those of previous works, and we discuss the caveats of our prescription.

\subsection{Comparison to previous works}

To compare the total formation efficiencies achieved in our clouds to those presented in previous works, we use the well-established trend of increasing formation efficiency in higher gas surface density environments \citep[e.g.][]{fall10,murray10,thompson16,grudic18b}. 

\begin{figure}
  \includegraphics[width=\hsize]{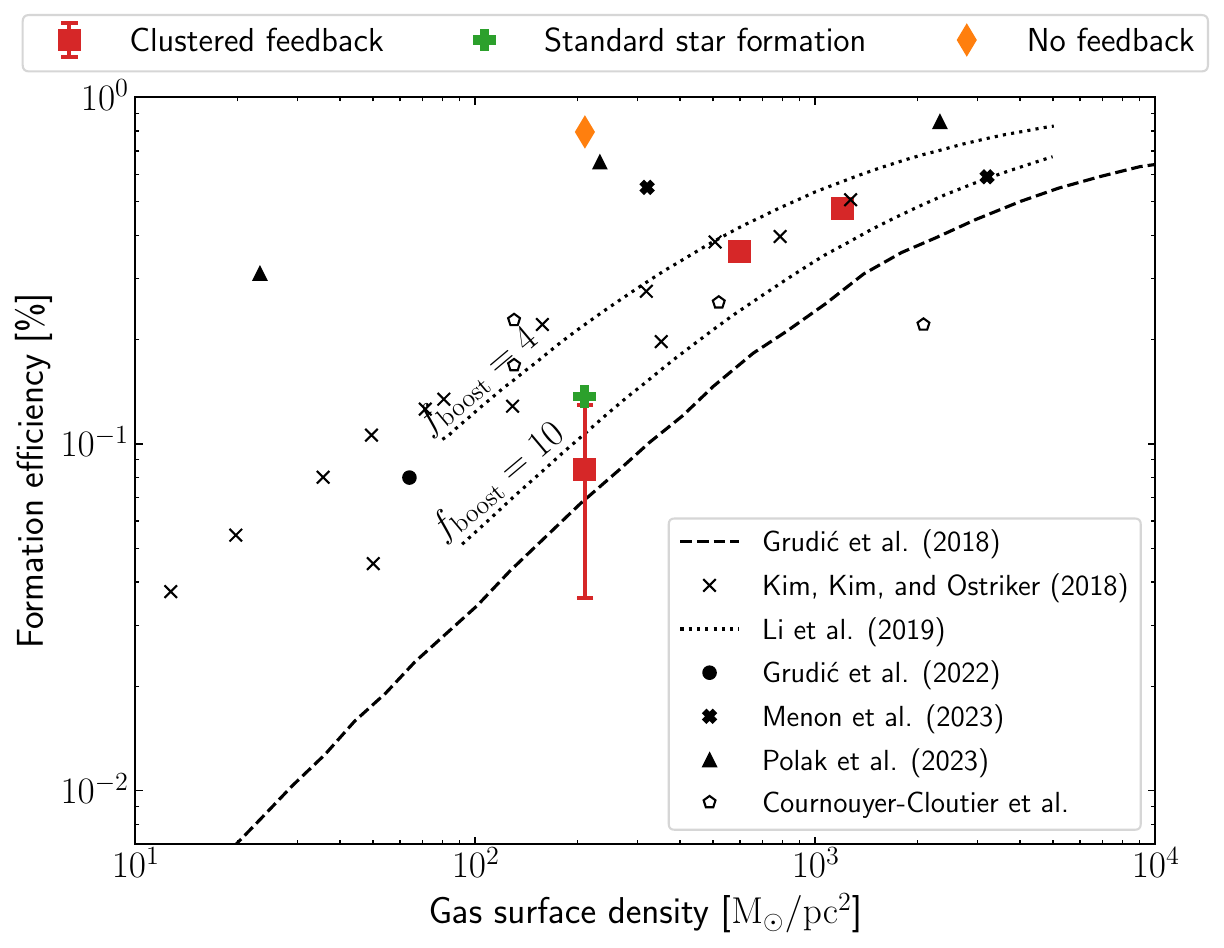}
  \caption{Comparison of the integrated star formation efficiency as a function of the initial gas surface density in our sample of clouds (coloured markers) against a variety of previous works (black lines and markers). The red errorbar shows the range of formation efficiencies in the clouds in which either the resolution, the gravitational softening and the sink formation criteria are changed, and the red marker corresponds to the fiducial cloud.}
  \label{fig:formEfficiency-int-comparison}
\end{figure}

In particular, we compare our results against previous simulations of isolated turbulent clouds that resolve single stars \citep[][and Cournoyer-Cloutier in prep.]{grudic22a,menon23,polak23} or low-mass stellar populations \citep[][]{grudic18b,kim18,li19b}. These works use different hydrodynamical codes (both Eulerian and Lagrangian), and make a myriad of assumptions for the stellar feedback mechanisms and their implementation. When possible, we use their measured formation efficiencies at $t=2t_{\rm ff}$ in Fig.~\ref{fig:formEfficiency-int-comparison}.

Despite using different modelling techniques for the star particles, we find that overall, the total formation efficiencies of our sample of clouds lie in very good agreement with these previous works. The only exception are the efficiencies reported by \citet{polak23}. Their low initial virial parameter ($\alpha_{\rm vir} = 0.15$) causes the cloud to collapse very quickly, producing similar efficiencies to our `No feedback' cloud. The clouds evolved with our clustered prescription lie around $\sim3$--$10~$per cent, in line with other results at a similar gas surface density.

\subsection{Caveats}\label{sub:caveats}

Our numerical prescription implicitly assumes that all star formation produces bound star clusters that do not dissolve over time. Although these assumptions are sufficient for the study of the formation of clusters in clouds under different choices as it is done here, expanding the scope to larger galactic environment will require addressing these caveats. In particular, future work will include a bound fraction within each sink (e.g. \citealt{li18} and the updated \citealt{brown22}), as well as prescriptions for the mass loss due to stellar evolution, relaxation, and tidal shocks (e.g.~sect.~3.2.1. in \citealt{reina-campos22b}, and \citealt{gieles23}). By modelling star clusters as individual particles within galaxies, the spiraling of massive clusters due to dynamical friction will be self-consistently accounted for.

\section{Conclusions}\label{sec:conclusions}

In this work, we present a numerical formalism to model star clusters as individual units of star formation within galaxies, which is implemented in the public version of the hydrodynamical code \gizmo \citep{hopkins15}. In our model, star clusters grow via gas accretion and merging with sub-clusters using a sink approach. Using a large grid of marginally bound, turbulent clouds, we explore the effect of different choices on the lifetime of the clouds and the formation of star clusters. 

Overall, in terms of the total formation efficiencies and the resulting cluster mass functions, we find that most of the tests follow similar evolution. Their total formation efficiencies are $\epsilon_{*}\sim10~$per cent, half of the star cluster mass has formed after $\sim1.5t_{\rm ff}$, star clusters assemble over $\sim0.3$--$0.6~\myr$ and their mass functions are well fit by a power-law of slope $\alpha\sim1.7$--$2.3$. 

There are two set of clouds that differ from this result: those evolved without radiative feedback, and those with higher gas surface densities. The presence of ionising radiation heats up the gas surrounding star clusters, which decreases the gas accretion rate predicted by the Bondi-Hoyle-Lyttleton rate and halts the growth of star clusters. Without this feedback mechanism, star clusters keep feeding from their natal gas, leading to total formation efficiencies of $\epsilon_* > 0.50~$per cent and shallower mass functions with slopes $\alpha \simeq 1.0$ \citep{rathjen21,andersson24}. In the clouds with increasing initial gas surface density, stellar feedback is less effective at halting the collapse of the gas \citep[e.g.][]{grudic18b}, which produces higher total formation efficiencies and more clusters that tend to be more massive.
 
Reassuringly, the fiducial cloud produces a similar formation efficiency of $\epsilon_*\sim 10~$per cent to the cloud in which the sinks are not allowed to accrete or merge, which we denote as `standard star formation'. These clouds evolve very similar until $t=1.7t_{\rm ff}$, when the ionising radiation emitted by the most massive cluster in the fiducial cloud quickly heats and ionises the entire cloud. We find that the feedback emitted by the clusters in the fiducial cloud is more efficient because it is injected in gas that tends to be at least an order of magnitude less dense than in the cloud with the standard star formation prescription. In the sink-based approach of the fiducial cloud, accretion converts dense gas particles that would otherwise keep collapsing into stellar populations, thus preventing runaway collapse towards high densities. In terms of the target resolution of future simulations, we find that a gas mass resolution in the range $100$--$1000~\msun$ is ideal to capture the detailed multiphase nature and structure of the ISM, and to produce a sufficiently wide mass range in the star cluster populations.  

The number of protocluster nurseries at early cosmic times is quickly increasing with the \textit{James Webb Space Telescope} \citep[e.g.][]{adamo24,mowla24}, and more information is available about their galactic environments \citep[][]{bradley24}. The formation of these massive, compact star clusters so early after the Big Bang suggests that studying star clusters within their host galaxy over cosmic time is paramount to bridge the scales between star formation, stellar feedback and galaxy evolution. 

\begin{acknowledgments}
MRC thanks James Beattie, Mike Grudić, Chris Matzner, and Vadim Semenov for useful conversations that improved this paper.

MRC gratefully acknowledges the Canadian Institute for Theoretical Astrophysics (CITA) National Fellowship for partial support; this work was supported by the Natural Sciences and Engineering Research Council of Canada (NSERC) [funding reference number 568580].
OG was supported in part by the U.S. National Science Foundation through grant AST-1909063 and by National Aeronautics and Space Administration through contract NAS5-26555 for Space Telescope Science Institute program HST-AR-16614.
HL is supported by the National Key R\&D Program of China No. 2023YFB3002502, the National Natural Science Foundation of China under No. 12373006, and the China Manned Space Program through its Space Application System.

The simulations were run in the Graham supercomputing cluster from Compute Ontario. The research was enabled in part by support provided by Compute Ontario (\href{https://www.computeontario.ca}{https://www.computeontario.ca}) and the Digital Research Alliance of Canada (\href{https://alliancecan.ca}{https://alliancecan.ca}).

\end{acknowledgments}

%

\vspace{5mm}
\facilities{Digital Alliance Canada - Graham}


\software{h5py \citep{h5py_allversions},
          Jupyter Notebooks \citep{Kluyver16}, 
          Numpy \citep{Harris20},
          Pandas \citep{pandas_allversions},
          Scipy \citep{Jones01},
          Matplotlib \citep{Hunter07},
          pynbody \citep{pynbody},
          DBSCAN \citep{ester96},
          astropy - a community-developed core Python package for Astronomy \citep{astropy13,price-whelan18} \footnote{\href{http://www.astropy.org}{http://www.astropy.org}}
          }



\appendix

\section{Stellar feedback prescriptions from FIRE-3}\label{app:fb}

We provide here a brief description of the relevant equations used to model the different stellar mechanisms in our code from \citet{hopkins23}. The quantities below are corrected to be multiplied by the instantaneous stellar mass $M_{*} = M_{*}(t)$, and the metallicity is defined in terms of the iron abundance as $\hat{z}=10^{\rm [Fe/H]}$. 

\subsubsection{Radiative feedback}

The bolometric luminosities of the stellar populations are well described by the piece-wise light-to-mass ratio:
\be
\dfrac{L/M_{*}}{\lsun/\msun} = \begin{cases}
		a_{L, 1} &(t\leq t_{L,1}) \\
		a_{L, 1}(t/t_{L,1})^{\psi_{L,1}} &(t_{L,1}<t\leq t_{L,2}) \\
		a_{L, 2}(t/t_{L,2})^{\psi_{L,2}}f_{L,2} &(t_{L,2}<t) \\
	\end{cases}
\ee
where the slopes $\psi_{L,n}$ and $f_{L,2}$ are
\be
\begin{aligned}
\psi_{L,1} &= \dfrac{\ln(a_{L,2}/a_{L,1})}{\ln(t_{L,2}/t_{L,1})},\\
\psi_{L,2} &= -1.82\left[1-0.1\left\{1-0.073\ln\left(\dfrac{t}{t_{L,2}}\right)\right\}\ln\left(\dfrac{t}{t_{L,2}}\right)\right],\\
f_{L,2} &= 1+1.2\exp\left[ -\dfrac{(\ln(t/t_{L,3}))^2}{a_{L,3}^2} \right],\\
\end{aligned}
\ee
and the coefficients $(a_{L,1}, a_{L,2}, a_{L,3}) = (800, 1100\hat{z}^{-0.1}, 0.163)$, and the timescales $(t_{L,1}, t_{L,2}, t_{L,3}) = (1.2, 3.7, 1200)~\myr$. 

The bolometric luminosity is divided into five different broad bands: ionizing ($\lambda < 912~$\AA), far-UV ($912~$\AA$\,< \lambda < 1550~$\AA), near-UV ($1550~$\AA $\,< \lambda < 3660~$\AA), optical and near-IR ($3660~$\AA $\,< \lambda < 3\mu$), and mid- and far-IR ($\lambda>3\mu$). To estimate the amount of photon flux in each band $L_{\nu} = f_{\nu} L$, we apply the intrinsic bolometric corrections $f_{\nu}$ described in appendix A in \citet{hopkins18}. Relative to those equations, the correction for the ionizing flux $f_{\rm ion}$ is updated to be \citep{hopkins23}, 
\be
f_{\rm ion} = \begin{cases}
		f_{\rm ion,1}  &(t < t_{\rm ion,1}) \\
3		f_{\rm ion,1}(t/t_{\rm ion,1})^{-2.9}  &(t_{\rm ion,1}<t\leq t_{\rm ion,2}) \\
		0  &(t > t_{\rm ion,2}) \\
	\end{cases}
\ee
with $f_{\rm ion,1} = 0.5$ and $(t_{\rm ion,1}, t_{\rm ion,2}) = (3.5, 150)~\myr$. We use the algorithmic implementation of radiative feedback within GIZMO as described in appendix E in \citet{hopkins18}.

\subsubsection{OB and AGB winds}

The continuous mass-loss rate from OB and AGB stellar winds is given by the piece-wise function\footnote{This rate and two coefficients are different between the original and the published version of \citet{hopkins23}. The last term in the mass-loss rate in the published version is $+ a_{A,1} \left(t/t_{A}\right)^{1.6}\left\{\exp\left(-(t_A/t)^6\right)+\left[a_{ A,2}^{-1} + \left(t_A/t\right)^{2}\right]^{-1}\right\}$, and the coefficients have changed to $t_{A} = 800~\myr$, $a_{A,1} = 0.11$. This work was developed with the original version of \citet{hopkins23} as posted on the arXiv repository server.},
\begin{equation}
\begin{aligned}
\dfrac{\dot{M}_{\rm w}/M_{*}}{\gyr^{-1}} &= \begin{cases}
		a_{w,1} &(t \leq t_{w,1}) \\
		a_{w, 1}(t/t_{w,1})^{\psi_{w,1}} &(t_{w,1} < t\leq t_{w,2}) \\
		a_{w, 2}(t/t_{w,2})^{\psi_{w,2}} &(t_{w,2} < t\leq t_{w,3}) \\
		a_{w, 3}(t/t_{w,3})^{\psi_{w,3}} &(t_{w,3} < t) \\
\end{cases}\\
&+ a_{A,1} \left\{ \left[1+\left(t/t_{A}\right)^{1.1}\right]\left[1+a_{ A,2}\left(t/t_{A}\right)^{-1}\right] \right\}^{-1}
\end{aligned}
\end{equation}
where the slopes $\psi_{w,n}$, coefficients $a_{w,n}$, and the timescales $t_{w,n}$ are
\begin{equation*}
\begin{aligned}
\psi_{w,1} &= \ln(a_{w,2}/a_{w,1})/\ln(t_{w,2}/t_{w,1}),\\
\psi_{w,2} &= \ln(a_{w,3}/a_{w,2})/\ln(t_{w,3}/t_{w,2}),\\
(a_{w,1}, &a_{w,2}, a_{w,3}, a_{A,1}, a_{A,2}) = (3\hat{z}^{0.87}, 20\hat{z}^{0.45},0.6\hat{z}, 0.01, 0.01),\\
(t_{w,1}, &t_{w,2}, t_{w,3}, t_{A}) = (1.7, 4.0, 20, 1000)~\myr.\\
\end{aligned}
\end{equation*}

The velocity of the mass-loss at injection is fit by,
\begin{align}
    \dfrac{v_{\rm w, inj}}{\kms} = \hat{z}^{0.12}&\left[ \dfrac{3000}{1+(t/t_{v,1})^{2.5}} + \dfrac{600}{1+\hat{z}^3(t/t_{v,2})^{6}+11.2\hat{z}^{1.5}} + 30\right] 
\end{align}
with $(t_{v,1}, t_{v,2}) = (3, 50)~\myr$, which determines the input momentum, $\dot{M}_{\rm w}v_{\rm w, inj}$, and kinetic energy, $0.5\dot{M}_{\rm w}{v_{\rm w, inj}}^2$, fluxes of the winds.

\subsubsection{Core-collapse supernovae}

The rate of core-collapse or Type II supernovae is well-fit by,
\be
\dfrac{R_{\rm CC}/M_{*}}{\gyr^{-1}\msun^{-1}} = \begin{cases}
		0 &(t < t_{S,1} \,{\rm or}\, t > t_{S,3}) \\
		a_{S, 1}(t/t_{S,1})^{\psi_{S,1}} &(t_{S,1}\leq t\leq t_{S,2}) \\
		a_{S, 2}(t/t_{S,2})^{\psi_{S,2}} &(t_{S,2}\leq t\leq t_{S,3}) \\
	\end{cases}
\ee
where the slopes $\psi_{S,n}$ are 
\begin{equation*}
    \begin{aligned}
    \psi_{S,1} &= \ln(a_{S,2}/a_{S,1})/\ln(t_{S,2}/t_{S,1}),\\
    \psi_{S,2} &= \ln(a_{S,3}/a_{S,2})/\ln(t_{S,3}/t_{S,2}),
    \end{aligned}
\end{equation*}
the coefficients $(a_{S,1}, a_{S,2}, a_{S,3}) = (0.39, 0.51, 0.18)$, and the timescales $(t_{S,1}, t_{S,2}, t_{S,3}) = (3.7, 7.0, 44)~\myr$.

Given a rate of SNe, the number of SNe that actually explode over a certain timestep $\Delta t$ is determined as the integer part of $R_{\rm CC}\Delta t$, with the round-off decimals used to stochastically add events. Each SNe event releases an energy that is weakly dependent on the metallicity,
\be
E_{\rm CC} = \max\left[(\hat{z} + 10^{-4})^{-0.12}, 1\right] 10^{51}~\erg,
\ee
and total mass ejected per event is given by,
\be
M_{\rm CC, ej} = 10~\msun(t/6.5~\myr)^{-\psi_{\rm CCM}},
\ee
where the slope $\psi_{\rm CCM} = 2.22$ for $t\leq 6.5~\myr$ and $\psi_{\rm CCM} = 0.267$ for $t > 6.5~\myr$.

\subsubsection{Thermal supernovae}

The rate of Ia supernovae is given by,
\be
\dfrac{R_{\rm Ia}/M_{*}}{\gyr^{-1}\msun^{-1}} = \begin{cases}
		0 &(t < t_{\rm Ia,1}) \\
		a_{\rm Ia, 1}(t/t_{\rm Ia,1})^{\psi_{\rm Ia,1}} &(t_{\rm Ia,1}\leq t) \\
	\end{cases}
\ee
where the onset of SNIa is assumed to be at the time of the last core-collapse SNe, $t_{\rm Ia, 1} = t_{S, 3} = 44~\myr$, and the slope and coefficient are $\psi_{\rm Ia,1} = -1.1$ and $a_{\rm Ia, 1} = 0.0083$. This relation is not given by the results from STARBURST99, but instead taken from \citet{maoz17}. The initial kinetic energy of the ejecta is assumed to be a constant value of $E_{\rm Ia} = 10^{51}~\erg$ and all of these events eject $M_{\rm Ia} = 1.4~\msun$. As in the case of core-collapse SNe, the actual number of events over a timestep $\Delta t$ is taken to be the integer part of $R_{\rm Ia}\Delta t$, with the decimals being used to add events stochastically.

\subsection{Chemical yields}

In addition to the energy and mass emitted by winds, radiation, and supernovae, we also account for the yields ejected via these feedback mechanisms. We track the total metallicity, Z, as well as nine different chemical abundances: He, C, N, O, Ne, Mg, Si, S, Ca, Fe. To define the solar abundances, we follow \cite{hopkins23} and update them to be scaled to the \citet{asplund09} proto-solar abundances with mass fractions of (Z, He, C, N, O, Ne, Mg, Si, S, Ca, Fe)$_{\odot}$ = ($0.0142$, $0.2703$, $2.53\times 10^{-3}$, $7.41\times 10^{-4}$, $6.13\times 10^{-3}$, $1.34\times 10^{-3}$, $7.57\times 10^{-4}$, $7.12\times 10^{-4}$, $3.31\times 10^{-4}$, $6.87\times 10^{-5}$, $1.38\times 10^{-3}$) \citep[see also][]{lodders09}. The chemical yield $y_{j}$ for a species $j$ is defined as the dimensionless mass-fraction-of-ejecta.

\subsubsection{Stellar winds from OB and AGB}

Given a total mass-loss rate, $f_{X}$ gives the mass fraction in a given species $X$. For heavy elements, negligible enrichment is predicted from outflows, so the abundance remains as the initial surface abundance, $f_{X} = f_{X, 0}$. However, for the species tracked, models predict more complex returns for He, N, C, and O. \cite{hopkins23} model the evolution of these abundances in terms of the processes leading to the creation and destruction of each species,
\be
\left\{
\begin{aligned}
    {\rm For\,He},\,f_{\rm He} &= f_{\rm He, 0}(1-y_{\rm HeC}) + y_{\rm HHe}f_{\rm H,0}, \\
    {\rm For\,N},\,f_{\rm N} &= f_{\rm N, 0} + y_{\rm CN}f_{\rm C,0}+ y_{\rm ON}f_{\rm O,0}, \\
    {\rm For\,C},\,f_{\rm C} &= f_{\rm C, 0}(1-y_{\rm CN}) + y_{\rm HeC}f_{\rm He,0} \\
        &+ y_{\rm HC}f_{\rm H,0}(1-y_{\rm HHe}), \\
    {\rm For\,O},\,f_{\rm O} &= f_{\rm O, 0}(1-y_{\rm ON}).
\end{aligned}
\right.
\ee
The different terms in the abundances above are $y_{\rm HeC} = y_{\rm HC}$, $y_{\rm CN} = \min\left[1, 0.5y_{\rm CNO} (1+x_{\rm OC})\right]$, $y_{\rm ON} = y_{\rm CNO} + (y_{\rm CNO} - y_{\rm CN})x_{\rm OC}^{-1}$ and $x_{\rm OC} = f_{\rm O, 0}/f_{\rm C, 0}$. Thus, the yields are entirely determined by the initial abundances and the yields $y_{\rm HHe}$, $y_{\rm CNO}$, and $y_{\rm HC}$. These can be determined as piece-wise functions, 
\be
y_{j={\rm HHe, CNO, HC}} = 
\begin{cases}
    a_{j, 1}(t/t_{j, 1})^{\psi_{j, 0}} &(t \leq t_{j, 1}) \\
    a_{j, 1}(t/t_{j, 1})^{\psi_{j, 1}} &(t_{j, 1} < t \leq t_{j, 2}) \\
    ... \\
    a_{j, n}(t/t_{j, n})^{\psi_{j, n}} &(t_{j, n} < t \leq t_{j, n+1}) \\
\end{cases} \label{eq:yields-winds}
\ee
where the slopes are $\psi_{j, n} = \ln(a_{j, n+1}/a_{j, n})/\ln(t_{j, n+1}/t_{j, n})$ for $n>0$ and all the species-dependent coefficients $a_{j, n}$ and the time boundaries $t_{j, n}$ are given in Table~\ref{tab:coeff-yields-snia}. Since the CNO metallicity drives the dependence of the light-element fusion, these yields are parametrized in terms of $z_{\rm CNO} = (Z_{\rm C,0}+Z_{\rm N,0}+Z_{\rm O,0})/(Z_{\rm C}+Z_{\rm N}+Z_{\rm O})_{\odot}$.
\begin{table*}
\caption{Coefficients for stellar mass-loss yields described in eq.~(\ref{eq:yields-winds}).}\label{tab:coeff-yields-snia}
\centering{
\begin{tabular}{ccccccc}
  \hline
  Term & $x_{j,1}$ & $x_{j,2}$ & $x_{j,3}$ & $x_{j,4}$ & $x_{j,5}$ & $x_{j,6}$ \\ \hline
  HHe ($y_{
\rm HHe}$; $\psi_{{\rm HHe},0}=3$) & & & & & & \\
$t_{\text{HHe},n}$ [Gyr] & 0.0028 & 0.01 & 2.3 & 3.0 & 100 & – \\
  $a_{{\rm HHe}, n}$ & 0.4 MIN[$(z_{\text{CNO}} + 0.001)^{0.6}, 2$] & 0.08 & 0.07 & 0.042 & 0.042 & – \\ \hline
  CNO ($y_{\rm CNO}$; $\psi_{{\rm CNO},0}=3.5$) & & & & & & \\
$t_{{\rm CNO},n}$ [Gyr] & 0.001 & 0.0028 & 0.05 & 1.9 & 14 & 100 \\
  $a_{\rm CNO, n}$ & 0.2 MIN[$z_{{\rm CNO}}^2 + 10^{-4}, 0.9$] & 0.68 MIN[$(z_{{\rm CNO}} + 0.001)^{0.1}, 0.9$] & 0.4 & 0.23 & 0.065 & 0.065 \\ \hline
  HC ($y_{\rm HC}$; $\psi_{{\rm HC},0}=3$) & & & & & & \\
  $t_{{\rm HC},n}$ [Gyr] & 0.005 & 0.04 & 10 & 100 & – & – \\
  $a_{{\rm HC}, n}$ & $10^{-6}$ & 0.001 & 0.005 & 0.005 & – & – \\ \hline
\end{tabular}}
\end{table*}

\subsubsection{Core-collapse supernovae}

The ejected metal yields $y_{{\rm cc}, j}$ for the species j are described in terms of a piece-wise function, 
\be
y_{{\rm cc}, j} = 
\begin{cases}
    0 &(t \leq t_{{\rm cc}, 1}) \\
    a_{{\rm cc}, j, 1}(t/t_{{\rm cc}, 1})^{\psi_{{\rm cc}, j, 1}} &(t_{{\rm cc}, 1} < t \leq t_{{\rm cc}, 2}) \\
    ... \\
    a_{{\rm cc}, j, n}(t/t_{{\rm cc}, n})^{\psi_{{\rm cc}, j, n}} &(t_{{\rm cc}, n} < t \leq t_{{\rm cc}, n+1}) \\
    0 &(t \geq t_{{\rm cc}, n+1}) \\
\end{cases}\label{eq:yields-cc}
\ee
where the slopes are
\begin{equation*}
    \psi_{{\rm cc}, j, n} = \ln(a_{{\rm cc}, j, n+1}/a_{{\rm cc}, j, n})/\ln(t_{{\rm cc}, n+1}/t_{{\rm cc}, n}),
\end{equation*} and all the species-dependent coefficients $a_{{\rm cc}, j, n}$ and the time boundaries $t_{{\rm cc}, n}$ are given in Table~\ref{tab:coeff-yields-sne-cc}.

\begin{table}
\caption{Coefficients for the yields ejected by core-collapse supernovae described in eq.~(\ref{eq:yields-cc}).}\label{tab:coeff-yields-sne-cc}
\centering{
\begin{tabular}{cccccc}
  \hline
  Species & $a_{{\rm cc}, j,1}$ & $a_{{\rm cc}, j,2}$ & $a_{{\rm cc}, j,3}$ & $a_{{\rm cc}, j,4}$ & $a_{{\rm cc}, j,5}$ \\ \hline
  He & 0.461 & 0.330 & 0.358 & 0.365 & 0.359 \\
  C & 0.237 & 8.57e-3 & 1.69e-2 & 9.33e-3 & 4.47e-3 \\
  N & 1.07e-2 & 3.48e-3 & 3.44e-4 & 3.72e-3 & 3.50e-3 \\
  O & 9.53e-2 & 0.102 & 9.85e-2 & 1.73e-2 & 8.20e-3 \\
  Ne & 2.60e-2 & 2.20e-2 & 1.93e-2 & 2.70e-3 & 2.75e-3 \\
  Mg & 2.89e-2 & 1.25e-2 & 5.77e-3 & 1.03e-3 & 1.03e-3 \\
  Si & 4.12e-4 & 7.69e-3 & 8.73e-3 & 2.23e-3 & 1.18e-3 \\
  S & 3.63e-4 & 5.61e-3 & 5.49e-3 & 1.26e-3 & 5.75e-4 \\
  Ca & 4.28e-5 & 3.21e-4 & 6.00e-4 & 1.84e-4 & 9.64e-5 \\
  Fe & 5.46e-4 & 2.18e-3 & 1.08e-2 & 4.57e-3 & 1.83e-3 \\
\hline
\multicolumn{5}{c}{Time boundaries $t_{{\rm cc}, n}$ [Myr]}\\\hline
  - & $t_{{\rm cc}, 1}$ & $t_{{\rm cc}, 2}$ & $t_{{\rm cc}, 3}$ & $t_{{\rm cc}, 4}$ & $t_{{\rm cc}, 5}$ \\
  - & $3.7$ & $8$ & $18$ & $30$ & $44$ \\
\end{tabular}}
\end{table}

\subsubsection{Thermal supernovae}

The chemical yields ejected by Type Ia SNe are taken from averaging the W7 and WDD2 models from \citet{leung18}: ($y_{\rm Z}$, $y_{\rm H}$, $y_{\rm He}$, $y_{\rm C}$, $y_{\rm N}$, $y_{\rm O}$, $y_{\rm Ne}$, $y_{\rm Mg}$, $y_{\rm Si}$, $y_{\rm S}$, $y_{\rm Ca}$, and $y_{\rm Fe}$) = ($1$, $0$, $0$, $1.76\times 10^{-2}$, $2.10\times 10^{-6}$, $7.36\times 10^{-2}$, $2.02\times 10^{-3}$, $6.21\times 10^{-3}$, $0.146$, $7.62\times 10^{-2}$, $1.29\times 10^{-2}$, and $0.558$).

\section{Numerical tests}\label{app:numerical-tests}

\begin{figure*}
  \includegraphics[width=\hsize]{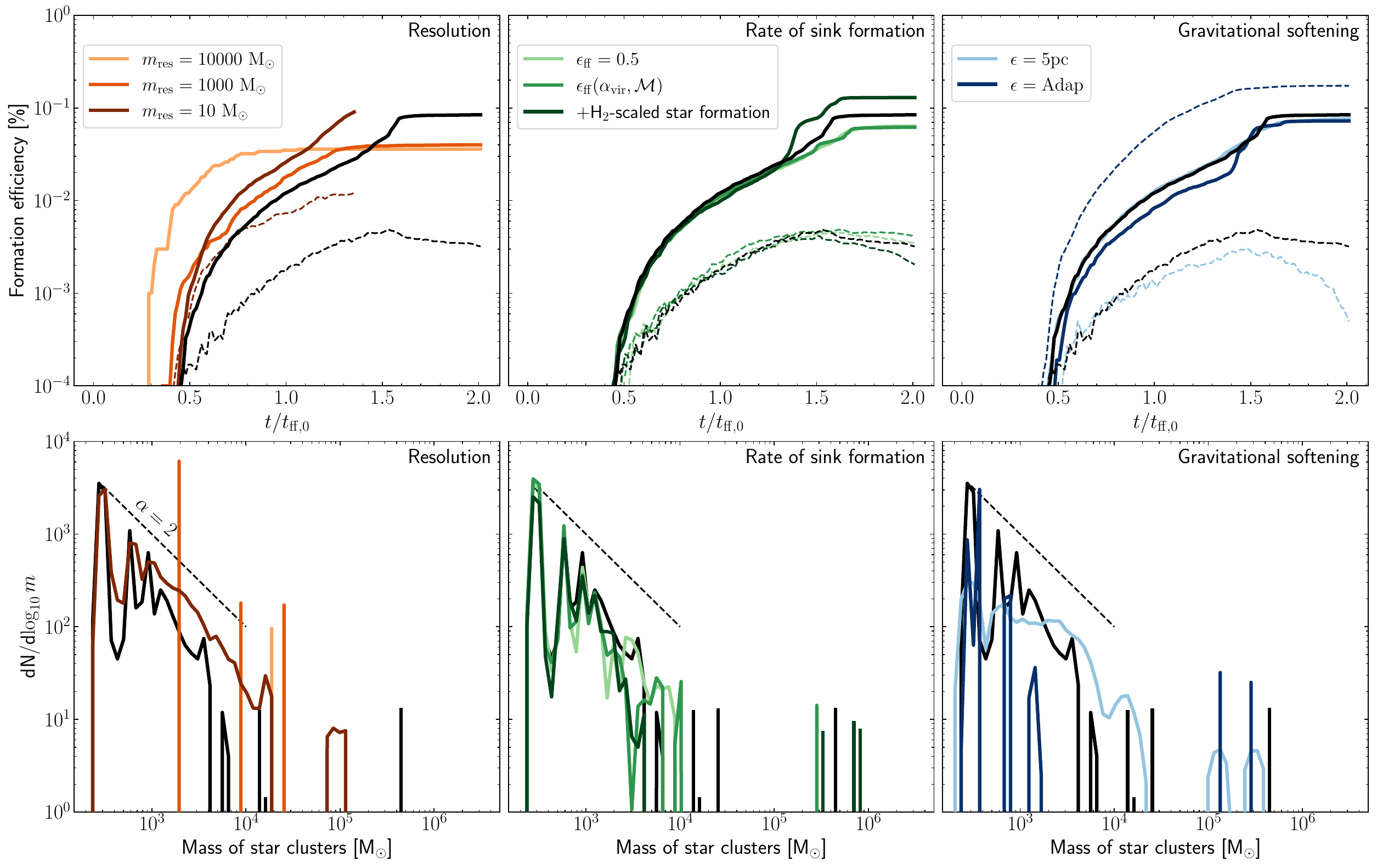}
  \caption{(\textit{Top:}) Formation efficiency of star clusters (thick solid line) and of sinks that remain starless (dashed thin line) amongst our numerical tests as a function of the initial free-fall timescale of the cloud. The black solid line present in all the panels corresponds to our fiducial cloud.
  (\textit{Bottom:}) Mass function of star clusters in our numerical tests calculated as a KDE with an Epanechnikov kernel. The thin dashed lines provide a visual representation of ${\rm d}N/{\rm d}m \propto m^{-2}$ between $300$--$10^4~\msun$. }
  \label{fig:app-formationEfficiency-dNdlog10m}
\end{figure*}

We describe here the results of the purely numerical tests: those in which we change the mass resolution of the gas particles, the rate at which sinks form, and the gravitational softening of the sinks. We show the time evolution of the formation efficiencies, and the resulting cluster mass functions in Fig.~\ref{fig:app-formationEfficiency-dNdlog10m}.

Changing the mass resolution of the cloud leads to similar total formation efficiencies: $\epsilon_{*} = 0.04$ for the two lowest mass resolutions ($m_{\rm gas} = 10^3, 10^4~\msun$), and $\epsilon_{*} = 0.09$ for the fiducial mass resolution of $m_{\rm gas} = 100~\msun$. The highest resolution cloud has formed $9\%$ of the initial cloud mass into star clusters by $t=10~\myr$, and that quantity would presumably continue to grow. Despite forming similar numbers of clusters, the clouds with the lowest resolution evolve faster, and formation stops after $\sim 5~\myr$ of evolution. Despite the similar efficiencies, the mass functions are affected by the different resolutions. Due to the stochastic accretion of entire gas cells into the sinks, clusters can never have masses smaller than the resolution element or that are not integer multiple of it. We thus recommend a maximum resolution of $500~\msun$ in order to be able to reproduce the mass function of young star clusters. Similarly, changing the criterion for the formation of the sinks only slightly changes the formation efficiency of star clusters or the slope of their mass function.

Assuming a larger fixed or adaptive gravitational softening for the sinks does not substantially alter the amount of mass that forms into clusters. However, the cloud evolved with an adaptive gravitational softening forms close to $20\%$ of its initial mass in sinks that remain starless. This indicates that a large number of sinks are prevented from merging, which would increase their mass and their accretion rate. In this test, we set the adaptive scale to be initially equal to that of the gas with the lowest gravitational softening being $1.23\times10^{-2}~\pc$. But the actual softening at the end of the simulation ranges between $1.5$--$122~\pc$, with a median of $30~\pc$, which lead to underestimated gravitational interactions between the sinks. This over-softened interactions between the clusters make them less likely to merge and grow in mass. Using a fixed softening larger than the fiducial value leads to a slightly shallower mass function with a slope of $\alpha = 1.6\pm 0.1$. 

\bibliography{mybib_cited}{}
\bibliographystyle{aasjournal}



\end{document}